\documentclass[secnumarabic,twocolumn,graphics,floatfix,nofootinbib,
tightenlines,nobibnotes,aps,prl]{revtex4-1}

\usepackage{subfigure}
\usepackage{color,amsmath,graphicx,epsfig,amssymb,picture}
\usepackage[T1]{fontenc}
\usepackage{pslatex}
\usepackage{txfonts}

\usepackage[svgnames]{xcolor}
\usepackage{hyperref}
\hypersetup{colorlinks,breaklinks,
            urlcolor=Blue,
            linkcolor=Blue,citecolor=Blue}

\begin{document}

\def\pd#1#2{\frac{\partial #1}{\partial #2}}
\def\bk{{\bf k}}
\def\bx{{\bf x}}
\def\br{{\bf r}}
\def\grad{{\mbox{\boldmath$\nabla$\unboldmath}}}

\title{\bf Long-range Ordering of Topological Excitations in a Two-Dimensional \\
Superfluid Far From Equilibrium}
\author {Hayder Salman and Davide Maestrini}
\affiliation {School of Mathematics, University of East Anglia, Norwich Research
Park, Norwich, NR4 7TJ, UK}

\begin {abstract}
We study the relaxation of a 2D ultracold Bose-gas from a nonequilibrium initial state containing vortex excitations in experimentally realizable square and rectangular traps. 
We show that the subsystem of vortex gas excitations results in the spontaneous emergence of a coherent superfluid flow with a non-zero coarse-grained vorticity field. 
The streamfunction of this emergent quasi-classical 2D flow is governed by a Boltzmann-Poisson equation. 
This equation reveals that maximum entropy states of a neutral vortex gas that describe the spectral condensation of energy can be classified into types of flow depending on whether or not the flow spontaneously acquires angular momentum. Numerical simulations 
of a neutral point vortex model and a Bose gas governed by the 2D Gross-Pitaevskii equation in a square reveal that a large scale monopole flow field with net angular momentum emerges that is consistent with predictions of the Boltzmann-Poisson equation.
The results allow us to characterise the spectral energy condensate in a 2D quantum fluid
 that bears striking similarity with similar flows observed in experiments of 2D classical turbulence. 
By deforming the square into a rectangular region,
the resulting maximum entropy state switches to a
dipolar flow field with zero net angular momentum.
\end{abstract}
\maketitle

Rapid experimental advances that have been made to realize superfluidity in low dimensions have paved the way to study nonequilibrium phenomena in such systems. These include the scenario of non-equilibrium relaxation of a system driven out of equilibrium, the Kibble-Zurek mechanism for defect formation following a temperature quench, and nonequilibrium regimes arising from a quantum phase transition \cite{Chomaz2015, Corman2014,Weiler2008,Desbuquois2012,Lamporesi2013,Sadler2006}. More recent work has focussed on uncovering under what conditions a quasiclassical regime of turbulence can emerge in a 2D spinless Bose superfluid where the topological excitations correspond to vortices and antivortices \cite{Neely2013,Billam2014,Numasato2010,Stagg2015,Nowak2011,White2014,Kwon2014}. 
Indeed, even holographic duals of superfluid and classical turbulence have been proposed to help identify the emergence of the quasiclassical regime \cite{Adams2014,Chesler2013}.
Turbulence in 2D is particularly intriguing as it behaves essentially differently from 3D. Following the classical theory of Kraichnan, Leith, and Batchelor \cite{Batchelor1969,Leith1968,Kraichnan1967}, it is known that, if the system is forced at intermediate scales, the energy will favour an upscale cascade. An important consequence of the inverse energy cascade in 2D is the spontaneous formation of large scale structures in the flow \cite{Xia2011,Francois2013}. For this reason, 2D turbulence has been motivated by its relevance to the emergence of large scale flow structures in quasi-geostrophic flows such as the ocean and planetary atmospheres \cite{Boffetta2012,Tabeling2002}. In contrast, such structures are absent in 3D turbulence that favours a downscale transport of energy. The essential difference between 2D and 3D arises as a consequence of the presence of a quadratic invariant for an inviscid 2D fluid called the enstrophy. In addition to energy, the enstrophy acts to severely constrain the spectral energy transfer in the system. Therefore, to sustain the inverse energy cascade in 2D, a downscale cascade in enstrophy is also observed.

In \cite{Billam2014,Simula2014} evidence of the emergence of an inverse cascade in models of 2D superfluid turbulence was presented while such an inverse cascade was argued to be absent in \cite{Numasato2010,Chesler2013}. In fact, the emergence of the inverse cascade is likely to be sensitive to the particular parameter regimes under consideration.
The existence of an inverse energy cascade permits spectral condensation of energy to occur. This scenario is analogous to the situation involving the formation of a Bose-Einstein condensate in an atomic gas from a non-equilibrium initial state \cite{Berloff2002,Davis2002}. In this case, it has been demonstrated that a particle flux towards low wavenumbers results in the formation of a condensate in the matter wave field. It is now well understood that characterising the properties of the coherent condensate field and its interaction with the incoherent thermal excitations is essential for a complete description of the system.

In this work, we present an analogous theory in order to describe the emergent coherent large flow patterns associated with the spectral condensation of energy in a 2D quantum fluid.
When our results describing the spectral energy condensate are combined together with the theory of the inverse energy cascade, 
we are able to characterise the key qualitative features of 2D quantum turbulence thus providing a more complete description of this phenomena in a quantum fluid. The theory we present also allows us to identify how a continuum coarse-grained enstrophy density emerges in a quantum fluid. This helps resolve how a direct enstrophy cascade that is associated with the process of filamentation of the vorticity field, can be sustained in a quantum system despite the fact that individual vortices in such a system are discrete.

It is well established that topological excitations in 2D play a key role in characterising the low temperature states of matter in Bose superfluids. In particular, it is known that even though a fully ordered Bose-condensate may not emerge, topological order can restore superfluidity in 2D at finite temperature as explained by the theory of Berezinskii-Kosterlitz-Thouless (BKT) \cite{Berezinskii1962,Kosterlitz1973}. 
This topological ordering also arises in thin films of liquid Helium and superconductors, in Ferromagnetism in 2D, in Coulomb gases \cite{Jose2013}, and in ultracold Bose gases \cite{Hadzibabic2006,Hadzibabic2011}. For a spinless Bose gas, the state of the topological vortex excitations is given by their positions in physical space. In this case, it follows from Onsager's theory \cite{Onsager1949} that the gas of vortex excitations in a confined domain can admit negative temperature states. This results in another form of long-range topological ordering in the limit of zero temperature that is associated with the collective motion of like-signed vortices as has been observed in \cite{Yatsuyanagi2005}. This collective behaviour results in the emergence of an order parameter, the streamfunction of the coarse-grained superfluid flow that demarcates the onset of the quasiclassical regime.
The long-range Coulomb-like interaction that exists between the vortices precludes a well-defined thermodynamic limit as in the case for gravitational problems and systems of electrically charged particles. We find that the resulting streamfunction of the coherent flow field is, therefore, dependent on the shape of the enclosing domain $\mathcal{D}$.
These observations are directly relevant to experiments in 2D where finite-size effects arise from the confinement of the condensate within a trapping potential.

In contrast to \cite{White2012} where vortex clusters were created by a moving obstacle, in \cite{Simula2014}, the phenomena of vortex clustering in a Bose gas was seen to emerge from a random initial distribution of vortices. It was shown that the annihilation of vortex-antivortex pairs drives the quantized vortex gas into the negative temperature regime through the process of  evaporative heating. The final large scale mean flow was shown to correspond to a dipole. In \cite{Simula2014}, a circular geometry was modelled which is a rather special case as the rotational symmetry imposes conservation of angular momentum thus severely constraining the dynamics of the vortices.
In order to uncover the different types of coherent flow that can emerge from the condensation of energy, we will consider a Bose gas that is trapped in a square box potential. A box potential has recently been realized in \cite{Gaunt2013}. 

\section{Gross-Pitaevskii Model}

We will model a condensate that is effectively trapped in a square box potential with principal axes aligned along the $x$ and $y$ coordinate directions and with a tight harmonic oscillator trap assumed along the $z$-direction. To simulate the key effects induced by the shape of the trap, we will impose reflective boundary conditions on the condensate wavefunction $\phi(x,y,t)$. The time evolution of the wavefunction is then given by the 2D Gross-Pitaevskii (GP) equation
\begin{eqnarray}
i\hbar \phi_t = -\frac{\hbar^2}{2m}\nabla^2 \phi + g_{\mathrm{2D}} |\phi|^2 \phi, 
\label{eqn_GP}
\end{eqnarray}
where $\phi(x,y) = 1/(\sqrt{2\pi}a_z) \int \phi_{\mathrm{3D}}(x,y,z) \exp{(-(z/a_z)^2)}$ is the axially integrated wavefunction, $g_{\mathrm{2D}} = g\sqrt{m\omega_z/(2 \pi \hbar)}$ is the effective 2D interaction parameter with $g=4\pi \hbar^2 a/m$, $a$ is the s-wave scattering length, and $a_z = \sqrt{\hbar/m\omega_z}=1.136\mu m$.
Motivated by the experiments of \cite{Neely2013}, we will assume a $^{87}$Rb condensate with $N=2\times 10^6$ atoms and with $\omega_z = 2\pi \times 90$ Hz. For these parameters, the healing length at $z=0$ is equal to $l_h  = \hbar/\sqrt{2mg_{\mathrm{2D}}N/L^2} = 0.235\mu m$ where we have set the extent of the condensate to correspond to $L^2 \sim (72 \mu m)^2$. These parameters are consistent with \citep{Neely2013,Samson2013} and imply that our system can contain many well-separated vortices. We note that the assumed trapping frequency along the $z$-coordinate direction does not completely freeze out the dynamics in the transverse direction. However, since we are simulating the system close to $T=0$, our assumptions to integrate out the dependence on $z$ is justified. Moreover, since we are particularly interested in the vortex dynamics, the 2D assumption for the vortices is valid provided that Kelvin waves excited along the length of the vortex would be damped efficiently. The length scale of the longest Kelvin waves that can exist in our case relative to the healing length satisfy $a_z/l_h \approx 4.8$. Since these scales are of a similar order, Kelvin waves would be damped effectively and hence the 2D assumption remains valid for the vortices also.

We solved a non-dimensional form of the GP equation by scaling space, time, and the wavefunction as $x \rightarrow (512/L)x$, $t \rightarrow 2mL^2/(512^2\hbar) t $, $\phi \rightarrow \sqrt{N} \phi $ respectively. It follows that $\tilde{g} = 2m g_{\mathrm{2D}} N/\hbar^2 = 93367$. The time evolution of the wavefunction is then governed by the 2D Gross-Pitaevskii (GP) equation given by
\begin{eqnarray}
i \phi_t = -\nabla^2 \phi + \tilde{g} |\phi|^2 \phi.
\label{eqn_GP}
\end{eqnarray}

The initial vortices are imprinted onto the condensate wavefunction by adapting the expression for the velocity potential (phase field) of a periodic array of vortices as described in \cite{Billam2014}. As we are interested in reflecting (Neumann) boundary conditions, we take our vortices to lie within a square box that is equivalent to a 1/4 of a periodic cell. The remaining 3/4 of the cell contain image vortices that are added in order to satisfy the reflective boundary conditions of the square box (see e.g.\ \cite{Campbell1991}). The resulting expression for the phase field corresponding to a neutral vortex gas consisting of $N^v$ vortices can then be written as
\begin{eqnarray}
\!\!\! \!\!\!\!\!\! \!\!\!\!\!\! && \varphi(x,y) = \sum_{k=1}^{(N^v/2)} \sum_{m=-\infty}^{\infty} g(X,Y)-g(X-2\pi , Y)-g(X,Y-2\pi) \nonumber \\ 
&& \hspace{2.5cm} +g(X-2\pi, Y-2\pi), \nonumber \\
&& g(X,Y) = \mathrm{atan} \left[ \tanh \left( \frac{Y_k^-}{2} + m\pi \right) \tan \left( \frac{X_k^- -\pi}{2}\right) \right] \\
&& - \mathrm{atan} \left[ \tanh \left( \frac{Y_k^+}{2} + m\pi \right) \tan \left( \frac{X_k^+ -\pi}{2}\right) \right] + \pi \left[ H(X_k^+) - H(X_k^-)) \right], \nonumber 
\end{eqnarray}
where $X_k^+ = \pi(x-x_k^{v,+})/L_x$, $X_k^- = \pi(x-x_k^{v,-})/L_x$, $Y_k^+ = \pi(y-y_k^{v,+})/L_y$, $Y_k^- = \pi(y-y_k^{v,-})/L_y$, and superscripts $\pm$ denote vortices with positive/negative circulation. In practice, the rapidly convergent infinite sum over $m$ allows us to truncate the series so that $m \in \{-5,5\}$.
The wavefunction is then reconstructed from $\phi(x,y) = \exp(i\varphi(x,y))$. The phase field obtained from the above expression is then frozen in time but the density of the condensate is relaxed through integration of the GP equation in imaginary time. This produces the desired distribution of vortices and antivortices with the required density profile in the condensate wavefunction. Using this as our initial condition for the wavefunction in the GP equation, we then integrate Eq.\ (\ref{eqn_GP}) forward in time using a four stage Strang Splitting method with a timestep of $\Delta t = 0.1$ using a discrete cosine transform. Unless stated otherwise, an $x-y$ grid of $(513 \times 513)$ points was used for the simulations in the square. For simulations in the rectangle, a grid of $(769 \times 513)$ was used.


\section{Point Vortex Model}

To provide a framework for formulating a statistical theory of the quantised vortex gas in order to explain the emergent coherent flow, we will also adopt a point vortex approximation which is applicable when the intervortex separation is much greater than the healing length. Assuming a hard wall potential which is consistent with our use of reflective boundary conditions and which provide a good approximation to the square/rectangular well potential assumed in this work, we use the Hamiltonian (equivalently renormalised Energy, $E$) given by \cite{Campbell1991,Campbell1996}, $H = (1/2)\sum_{i=1}^{N^v} \gamma_i \int \psi({\bf r})\delta({\bf r}-{\bf r}_{i,v}) {\rm d} {\bf r} - \mathrm{SI}$. The streamfunction corresponding to the flow field induced by the point vortices is
\begin{eqnarray}
&& \psi({\bf r}) = \frac{\rho}{2\pi N^v} \sum_{j=1}^{N^v} \frac{\gamma_j}{2}  \left[ f(|x-x_j|,|y-y_j|)-f(L_x-x_j-x,|y-y_j|) \right. \nonumber \\
&& \left. -f(|x-x_j|,L_y-y_j-y)+f(L_x-x_j-x,L_y-y_j-y) \right], \nonumber \\
&& f(\br) = f(x,y) = \frac{2\pi}{\alpha} \left[ \frac{|y|}{L_y} \left( \frac{|y|}{L_y}-1 \right)+\frac{1}{6} \right] \label{eqn_streamfunction} \\
&& -\ln \left\{ \prod\limits_{j=-\infty}^{\infty} \left[1-2\cos \left(\frac{2\pi x}{L_x} \right) e^{-2\pi |j+y/L_y|/\alpha}
+e^{-4\pi |j+y/L_y|/\alpha} \right] \right\},
\nonumber
\end{eqnarray}
where $\gamma_j=\pm 1$ is the circulation, and $\alpha = L_x/L_y$ is the aspect ratio of the domain. The divergent contributions from the self-interaction energies ($\mathrm{SI}$) and corresponding to the first term in the above expression for the streamfunction when $i=j$ are subtracted from the finite contributions to the Hamiltonian. We have set $\rho=1$ for the superfluid density. Hamilton's equations governing the time evolution of the vortex positions, $\dot{x}_{k,v} = \gamma_k^{-1} {\partial H}/{\partial y_{k,v}}, \dot{y}_{k,v} = -\gamma_k^{-1} {\partial H}/{\partial x_{k,v}}$ are then integrated numerically using an adaptive 4th/5th stage Runge-Kutta-Fehlberg scheme with a maximum time-step of $\Delta t=0.0005$.

\begin{figure}[t]
\centering
\hspace{-0.5cm}
 \begin{minipage}[b]{0.49\textwidth}
    \centering
        \includegraphics[width=3.in]{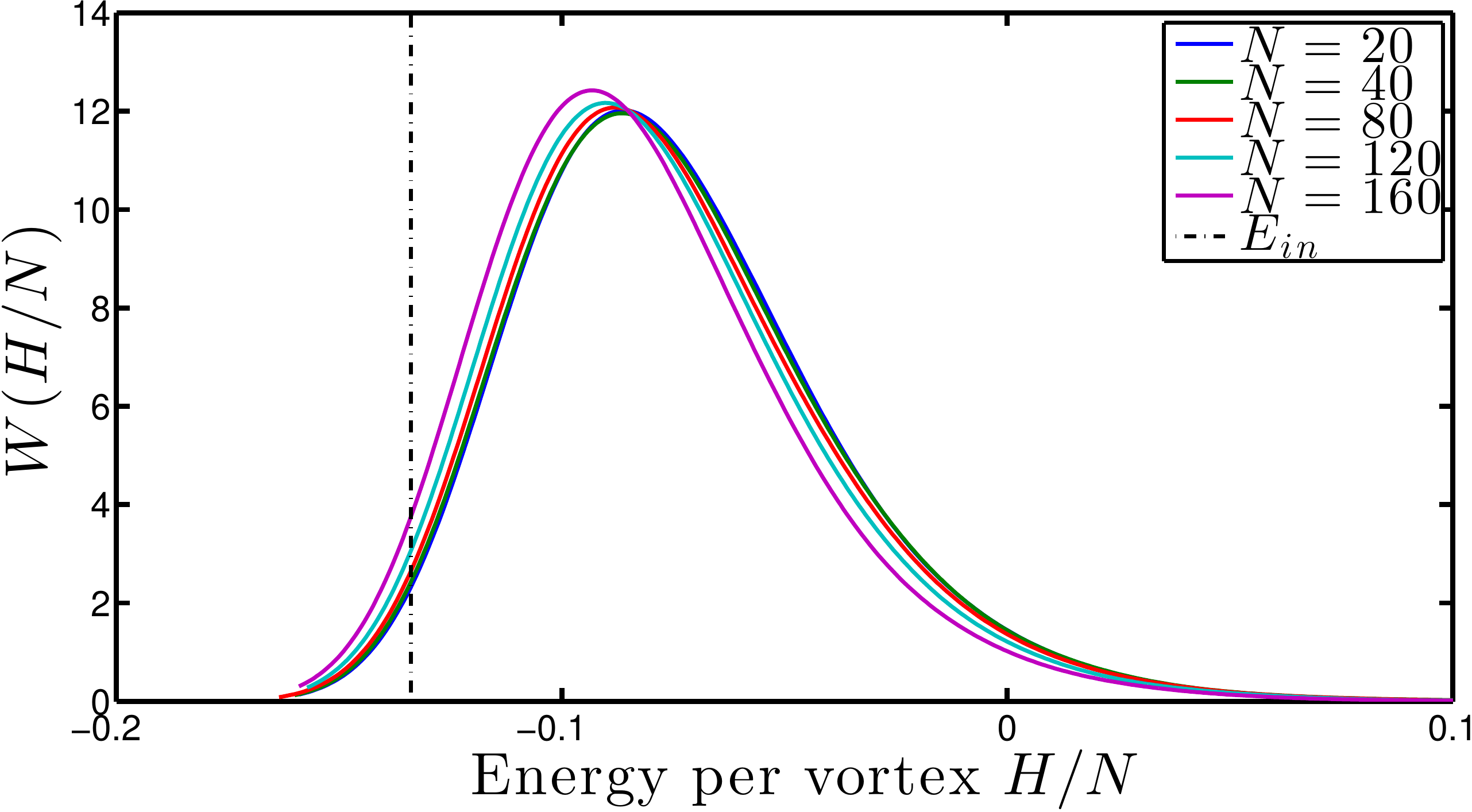}         
  \end{minipage}
 \vspace{-0.3cm}
  \caption{Statistical weight for vortices in square as a function of energy per vortex with miniumum intervortex and vortex-boundary separation of $0.12$ and $0.08$ imposed on vortex distributions. \label{fig_StatWeight}}
\vspace{-0.4cm}
\end{figure}

Using this Hamiltonian, we modelled vortices placed randomly with a uniform distribution within a domain of lengths $(L_x,L_y) = (2,2)$ in the $(x,y)$ coordinate directions, respectively. Constraints were imposed such that that the intervortex and the vortex-boundary separations were greater than $0.12$ and $0.08$, respectively. To ensure that our initial vortex distribution corresponds to a positive temperature, we calculated the statistical weights by sampling $M=100,000$ realizations of a neutral vortex gas consisting of $N^v$ vortices. For each realization, the interaction energy per vortex was calculated and a probability distribution constructed from \cite{Esler2013}
\begin{eqnarray}
W(H) = \frac{1}{M} \sum_i^{N_b} \frac{W_i}{\sigma\sqrt{2\pi}} e^{-\frac{(H-H_i)^2}{2\sigma^2}},
\end{eqnarray}
where $\sigma=0.01$, $W_i$ is the number of realizations with an energy within the interval $(H,H+\Delta H)$, and $N_b=50$ is the total number of bins. Fig.\ \ref{fig_StatWeight} presents the statistical weights for different $N^v$. We note that in comparison to the unconstrained case where the graphs converge as $N^v$ is increased \cite{Campbell1991}, our graphs do not converge since the constraint introduces a packing factor. However, in both cases, the qualitative features of the distribution persist. In particular, the probability distribution presented in Fig.\ \ref{fig_StatWeight} contains a maximum turning point with a positive slope at lower interaction energies and a negative slope at higher interaction energies. The entropy of the vortex gas is related to the statistical weight by $S=k_B \ln W$ with the temperature defined as $1/T = \partial S/ \partial E$. We, therefore, note that the system passes from positive to negative temperatures through $T=+\infty$. We chose our initial vortex distribution to lie within the positive temperature regime as indicated by the vertical line in Fig.\ \ref{fig_StatWeight}.

With these initial conditions, we integrated Hamilton's equations, for a neutral gas consisting of $N^v=120$ point vortices. Being a Hamiltonian system, the motion of the point vortices is constrained to lie on a surface of constant energy. To allow vortices to explore different regions of energy space, we model the mechanism of vortex-antivortex annihilation when two vortices approach each other within a separation distance that is twice the respective healing length as in \cite{Campbell1991,Simula2014}. The removal of vortex pairs results in a punctuated Hamiltonian model which breaks the invariants of the system at such vortex annihilation events. To model vortex-antivortex annihilation, we remove a vortex-antivortex pair when their intervortex separation falls below a critical value of $\delta = 0.01$. We note that $\delta/L \sim 0.005$ is in good agreement with $l_h/L § \sim 0.003$ for the parameters specified above for our GP simulations. Vortices in our GP simulations can also annihilate at the boundaries of the domain and thus change the polarization of the gas. However, we did not model this process in the point vortex model since our aim in using this model was to establish the validity of the statistical theory to be presented in the following section which is derived under the assumption of a neutral vortex gas.

\section{Mean Field Theory of Vortex Gas}

In this section, we present a statistical theory that we will adopt to explain the vortex distributions obtained from our numerical simulations that will be presented in the next section. Working in the microcanonical ensemble, we begin by dividing the flow domain into a large number of cells with area $\Delta \ll A$ where $A$ is the area of the domain. Each cell is assumed to contain a large number of point vortices. The macrostate is defined by the number of point vortices of species $a$ with circulation $\gamma_a$ in the $i$'th cell which we denote by the set $\{ N_{i,a}^v \}$. The statistical weight corresponding to the macrostate $\{ N_{i,a}^v \}$, is given by \cite{Joyce1973}
\begin{eqnarray}
\!\!\!\!\!\!\!\!\!\!\!\! W(\{ N_{i,a}^v \}) = \prod_a \left\{ N_{a}^v! \prod_i \frac{1}{N_{i,a}^v!} \left( \frac{\Delta}{A} \right)^{N_{i,a}^v} \right\}, \;\;\;\;\;\; N_{a}^v = \sum_i N_{i,a}^v.
\end{eqnarray}
By defining the coarse-grained vorticity field as
\begin{eqnarray}
\!\!\!\!\!\!\!\!\!\!\!\!   \omega(x,y) = \sum_a \omega_a(x,y) = \sum_a \gamma_a n_a(x,y), \;\;\;\; n_a(x_i,y_i) = N_{i,a}^v/\Delta^2,
\end{eqnarray}
and taking the continuum limit $\Delta \rightarrow 0$, we recover
\begin{eqnarray}
S = -k_B \sum_a \int_{\mathcal{D}} \frac{\omega_a}{\gamma_a} \log \left( \frac{\omega_a A}{\gamma_a N_a^v} \right) {\rm d}^2{\bf x}, 
\end{eqnarray}
for the entropy. We now maximise the entropy subject to the constraint of fixed energy, and fixed number of vortices and antivortices (i.e.\ we're considering the long time limit where the system has reached a quasi-equilibrium steady state such that vortex-antivortex annihilation can be neglected). This requires $\delta S - \beta \delta E - \sum_a \mu_a \gamma_a \delta n_a = 0$, where $\beta$ is the inverse temperature and $\mu_a$ is the respective chemical potential of each species. For a neutral quantized vortex gas ($\sum_a \gamma_a n_a = 0$), we recover the Boltzmann-Poisson equation \cite{Chavanis2014,Joyce1973}
\begin{eqnarray}
\nabla^2 \Psi + \frac{\lambda^2}{2} \left( \frac{\exp(\Psi)}{\int \exp(\Psi) {\rm d}^2{\bf x}} -\frac{\exp(-\Psi)}{\int \exp(-\Psi) {\rm d}^2{\bf x}} \right) = 0.
\label{eqn_BoltzPoiss}
\end{eqnarray}
Here $\lambda^2 = -N^v\gamma^2 \rho \beta/A$, $\Psi$ is the streamfunction of the mean flow satisfying $\nabla^2 \Psi = -\omega$, $\omega = \hat{\mathbf{z}} \cdot \nabla \times \mathbf{u}$ is the local (coarse-grained) vorticity field, and $\mathbf{u}  = \hat{\mathbf{z}} \times \grad \Psi$.
For positive temperatures where $\lambda^2<0$, Eq.\ (\ref{eqn_BoltzPoiss}) only has the trivial solution, $\Psi=0$. However, for negative temperatures, several solutions can be found for each value of $\lambda^2$, and hence we expect to have a non-trivial mean flow-field. We focus on states that are global maxima.

We have found solutions of this equation by adapting the method described in \cite{Book1975,McDonald1974} for the Sinh-Poisson equation to numerically solve Eq.\ (\ref{eqn_BoltzPoiss}). To enhance the stability of the method for the Boltzmann-Poisson equation, we have reexpressed the equation into the form
\begin{eqnarray}
&& \nabla^2 \Psi + \frac{\tilde{\lambda}^2}{2} \left( \sqrt{\frac{a^-}{a^+}} \exp(\Psi) -\sqrt{\frac{a^+}{a^-}} \exp(-\Psi) \right) = 0. \nonumber \\
&& a^+ = \frac{1}{A} \int \exp(\Psi) {\rm d}^2{\bf x}\, , \;\;\;\;\;\;\;\;\;\; a^- = \frac{1}{A} \int \exp(-\Psi) {\rm d}^2{\bf x}\, ,
\end{eqnarray}
and solved for prescribed values of $\tilde{\lambda}^2 = \lambda^2/\sqrt{a^+a^-}$.

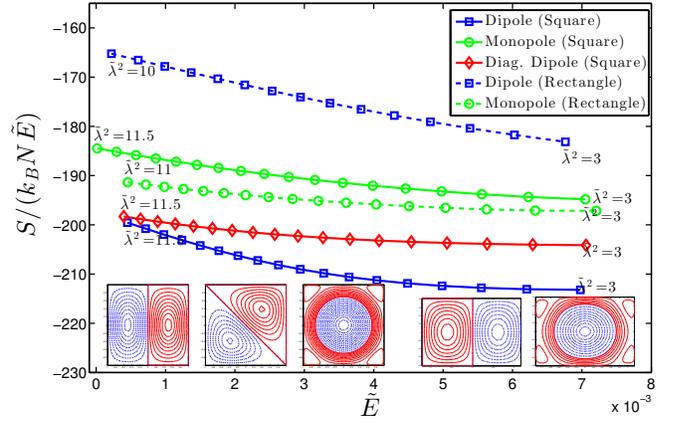
\begin{figure}[t]
\centering
\hspace{-0.75cm}
 \begin{minipage}[b]{0.52\textwidth}
    \centering
\sbox0{\includegraphics[width=3.372in]{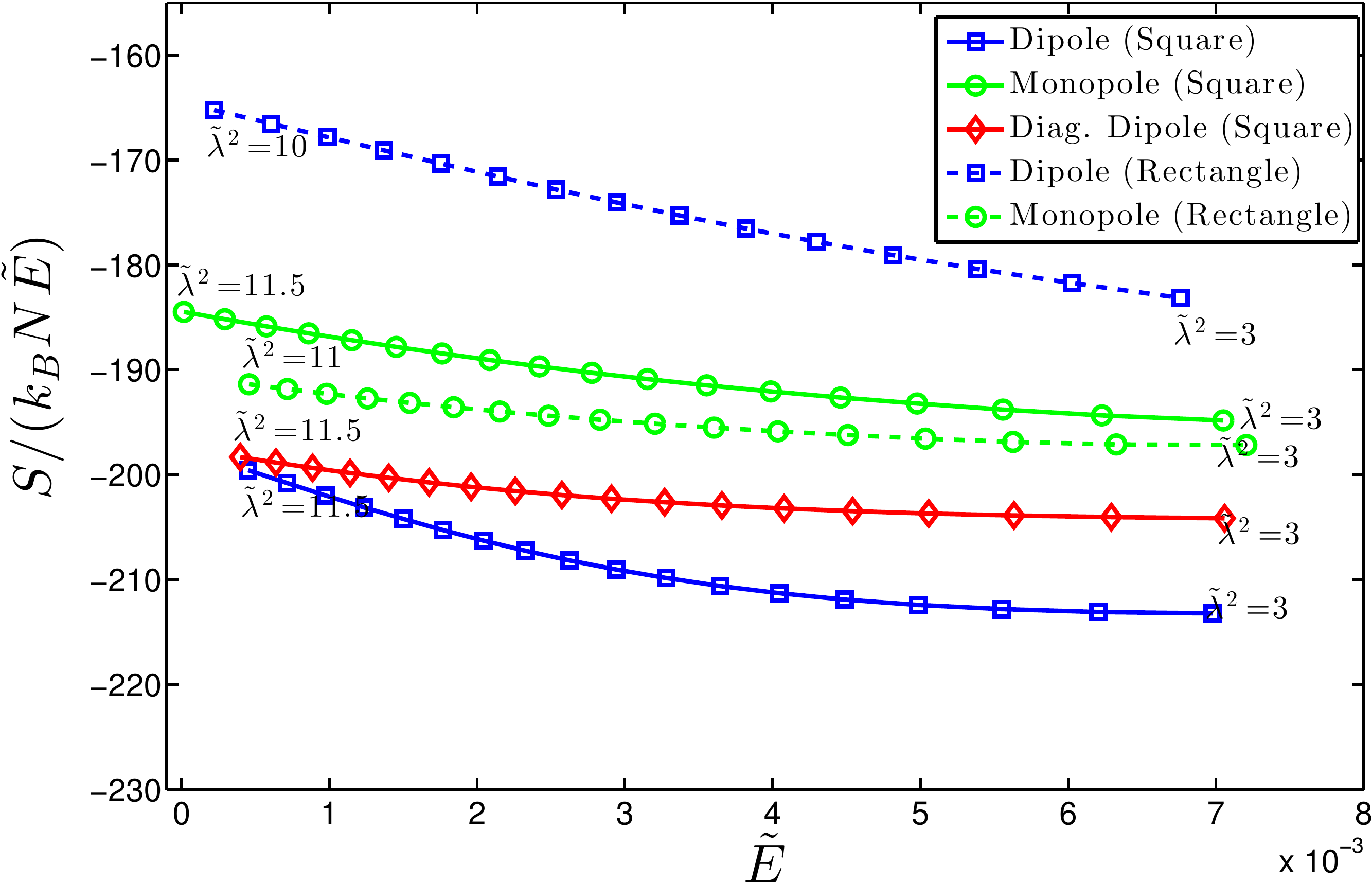}}
\sbox1{\includegraphics[width=0.45in]{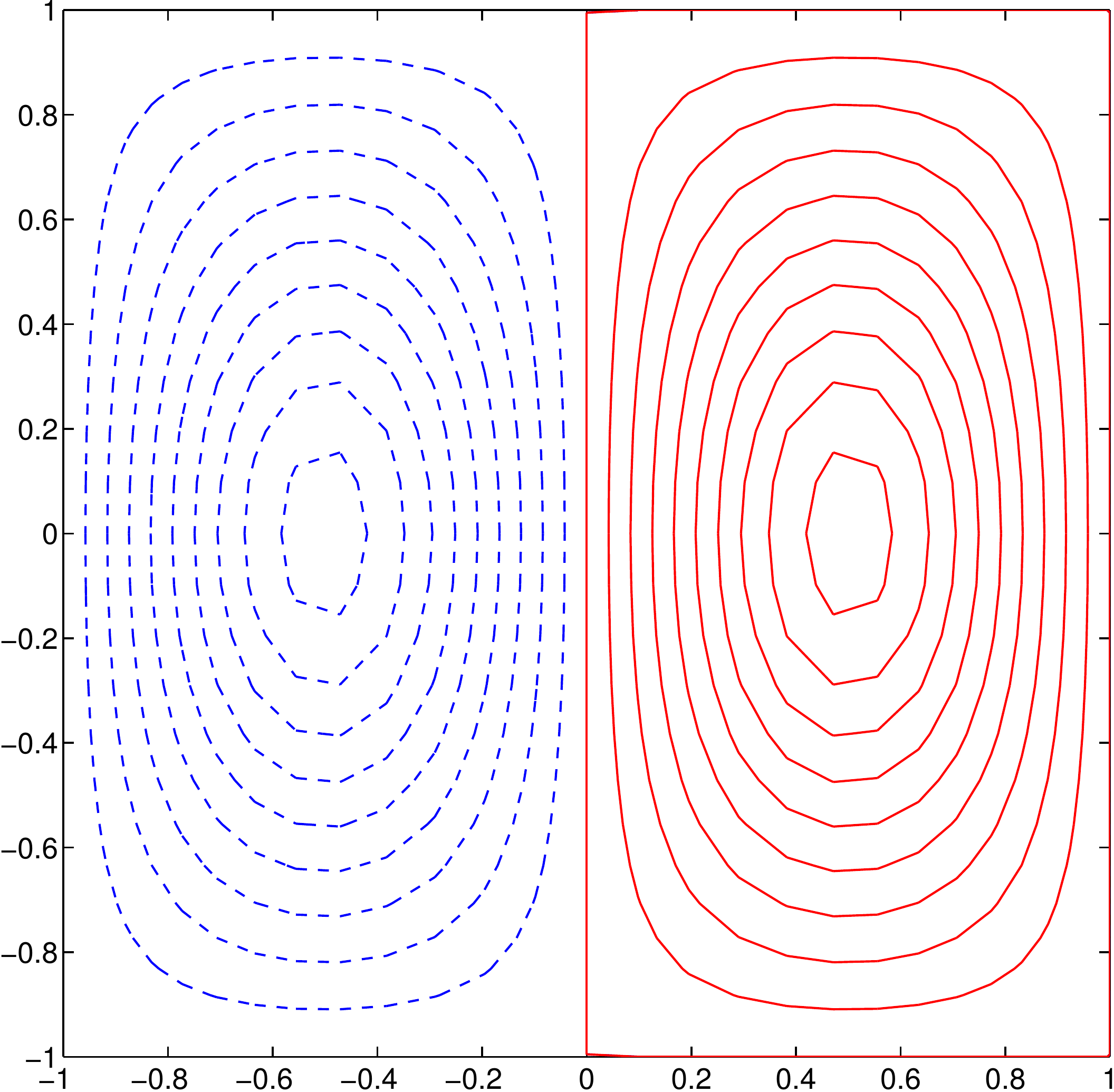}}
\sbox2{\includegraphics[width=0.45in]{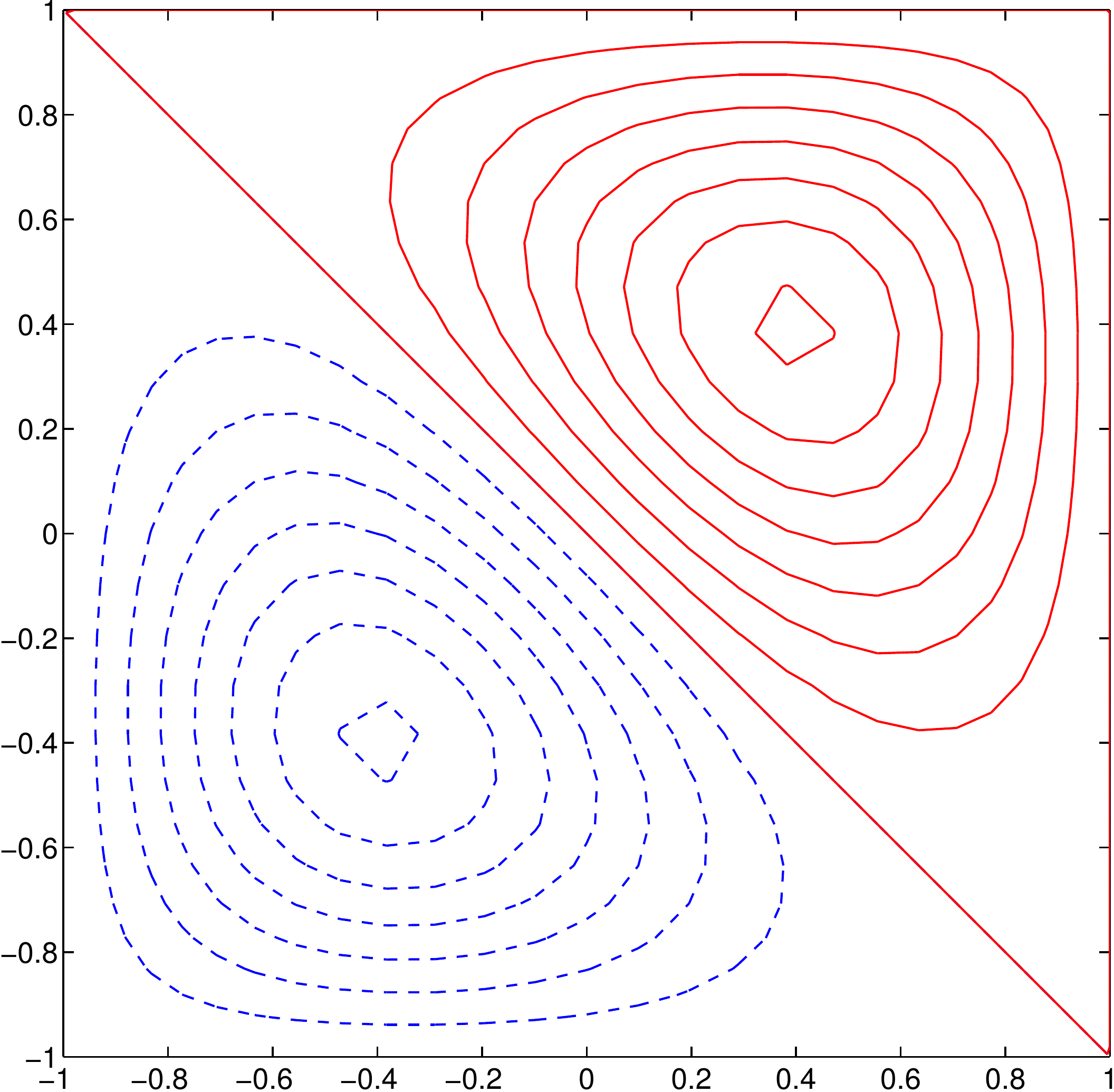}}
\sbox3{\includegraphics[width=0.45in]{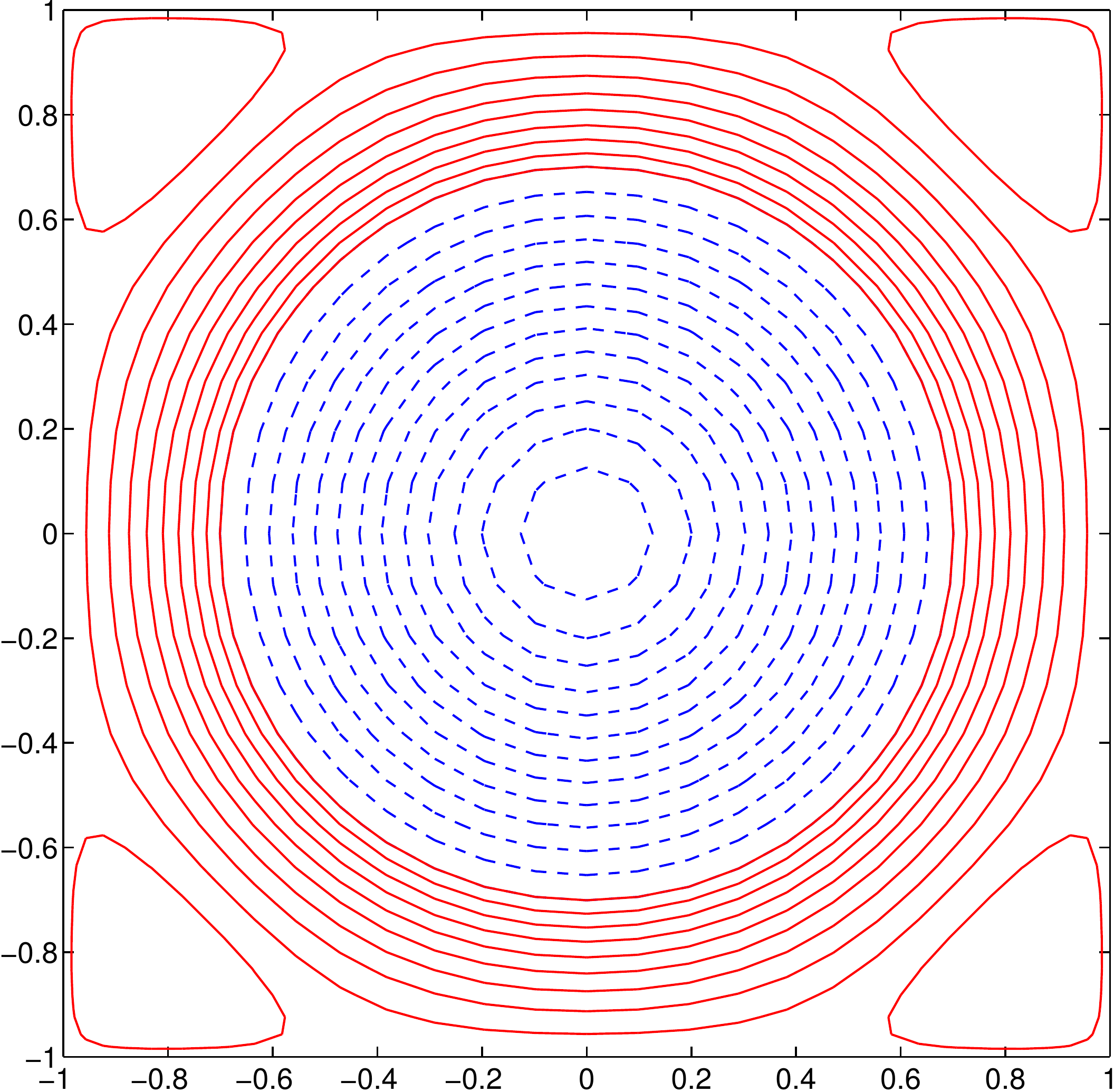}} 
\sbox4{\includegraphics[width=0.54in]{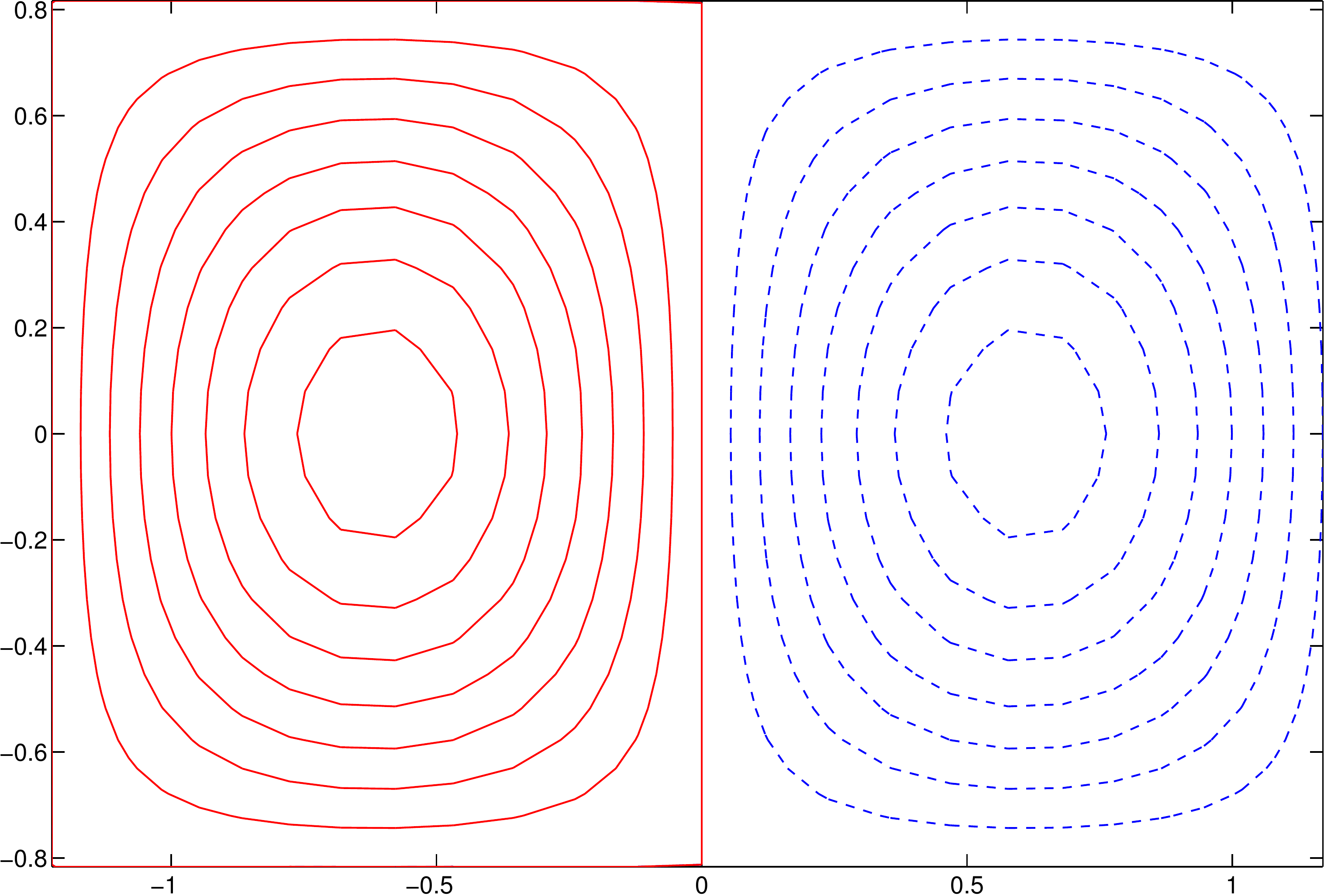}}
\sbox5{\includegraphics[width=0.54in]{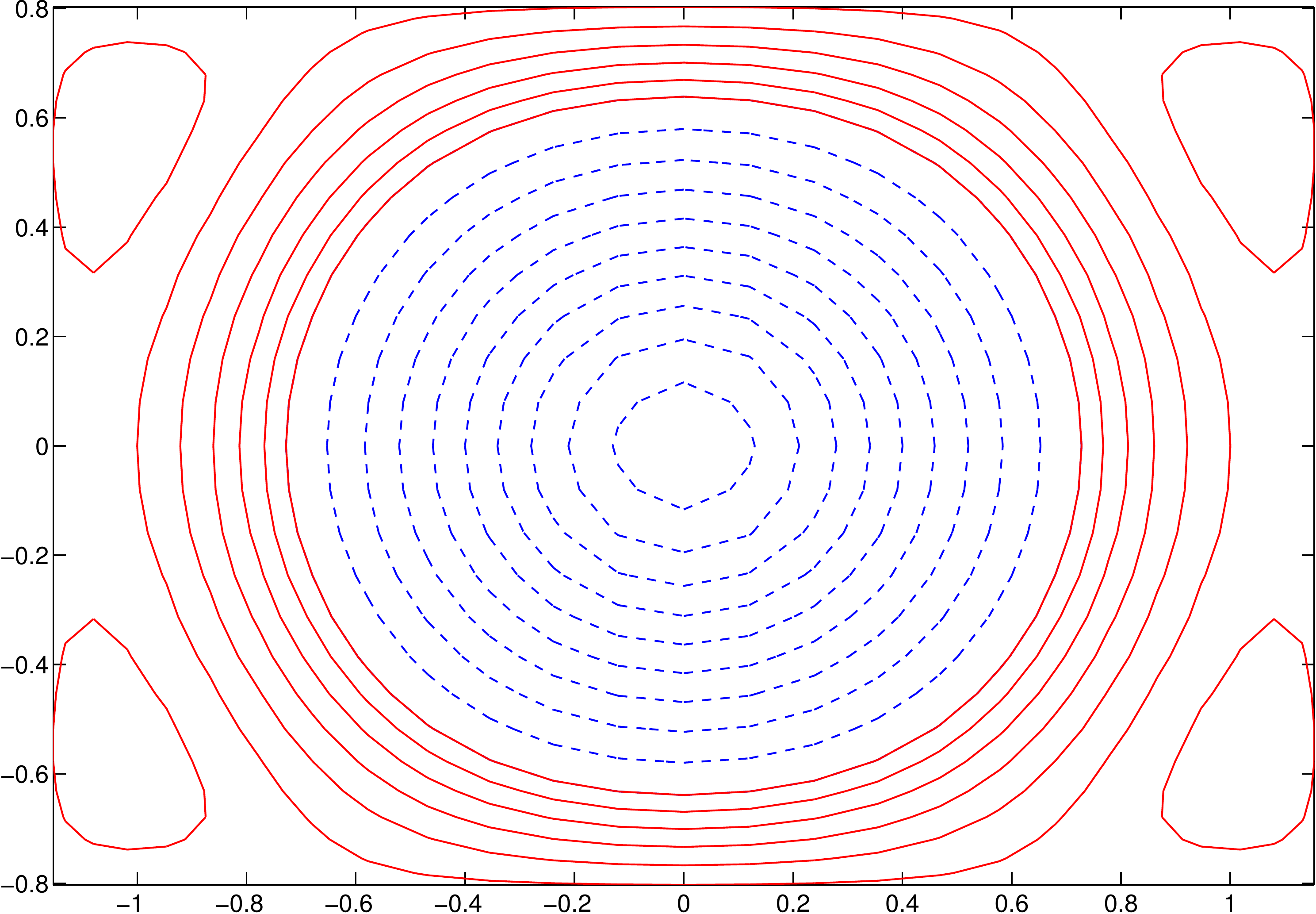}}
\begin{picture}(\wd0,\ht0)
\put(0,0){\usebox0}
\put(\wd0 - \wd2 - 6.2cm,\ht0 - \ht2 - 3.75cm){\usebox{1}}
\put(\wd0 - \wd2 - 4.9cm,\ht0 - \ht2 - 3.75cm){\usebox{2}}
\put(\wd0 - \wd2 - 3.6cm,\ht0 - \ht2 - 3.75cm){\usebox{3}}
\put(\wd0 - \wd2 - 2.0095cm,\ht0 - \ht2 - 3.75cm){\usebox{4}}
\put(\wd0 - \wd2 - 0.5cm,\ht0 - \ht2 - 3.75cm){\usebox{5}}
\end{picture}
\end{minipage}
 \vspace{-0.3cm}
  \caption{Entropy as a function of energy of mean flows predicted by Boltzmann-Poisson equation for square domain (dipole, diagonal dipole, and monopole solutions) and for rectangular domain (dipole, and monopole solutions). Insets show vorticity contours (red is postive and blue is negative) of mean flows.  \label{fig_BPbranches}}
\vspace{-0.4cm}
\end{figure}

In Fig.\ \ref{fig_BPbranches}, we present different solutions of $\Psi$ for the square and rectangular domains that are all local maximisers of the entropy. The variation of normalized entropy against normalized energy, where $\tilde{E} = \rho (8A^2\lambda^4)^{-1} \int \omega \Psi \mathrm{d}^2 \mathbf{x}$, is also presented for these different flows. We observe that, for the square, the monopole solution is a higher entropy state as discussed in \cite{Pointin1976,Taylor2009} and corresponds to a flow with a non-zero component of angular momentum $L_z= 2\rho \int \Psi \mathrm{d}^2 {\bf{x}}$. In contrast, the two dipolar flow fields correspond to lower entropy states with zero angular momentum.

We have seen that in contrast to the circular domain considered in \cite{Simula2014} which conserves angular momentum, the most probable mean flow in the square corresponds to a monopole. 
However, because of the long-range Coulomb-like interactions, the vortex gas does not have a well-defined thermodynamic limit. In fact, a linearized analysis of Eq\ (\ref{eqn_BoltzPoiss}) reveals that the branches corresponding to the monopole and dipole configurations will cross each other for a rectangle with an aspect ratio of $\alpha \simeq 1.12$ \cite{Taylor2009, Lundgren1977,Chavanis1996}.
Since analytical solutions exist only for some of these flows \cite{Ting1984,Ting1987}, we have numerically analysed the full nonlinear problem given by Eq.\ (\ref{eqn_BoltzPoiss}) by studying how these flow configurations change for a rectangle with aspect ratio $\alpha = 1.5$ with dimensions $(L_x,L_y)=(2\sqrt{\alpha},2/\sqrt{\alpha})$. Fig.\ \ref{fig_BPbranches} shows that for the rectangle, the relative position of the diagonal and monopole branches switches. Moreover, we have found that the two dipole branches coalesce into one. 

\section{Numerical Results of Point Vortex and Gross-Pitaevskii Simulations}

\begin{figure}[t]
\centering
 \begin{minipage}[b]{0.49\textwidth}
    \centering
    \subfigure[\label{fig_t0} \hspace{0.3cm}  $t = 0$, \hspace{1.25cm} $t=47.6t_v$, \hspace{1.35cm} $t = 65.97t_v$ \hfill]{
      \label{fig:mini:subfig:a}
      \includegraphics[width=0.85in]{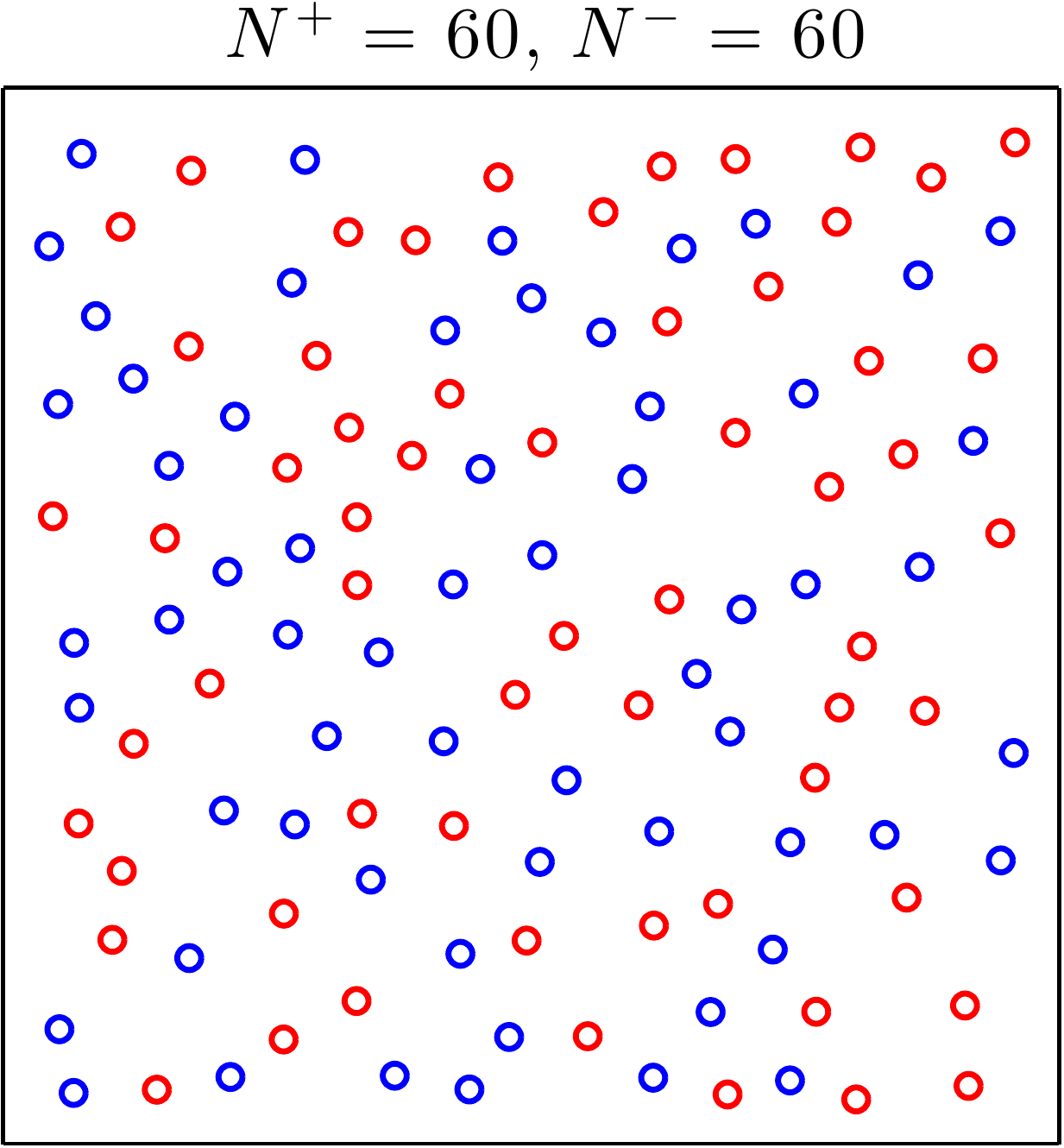}      
      \hspace{0.05cm}
      \includegraphics[width=0.85in]{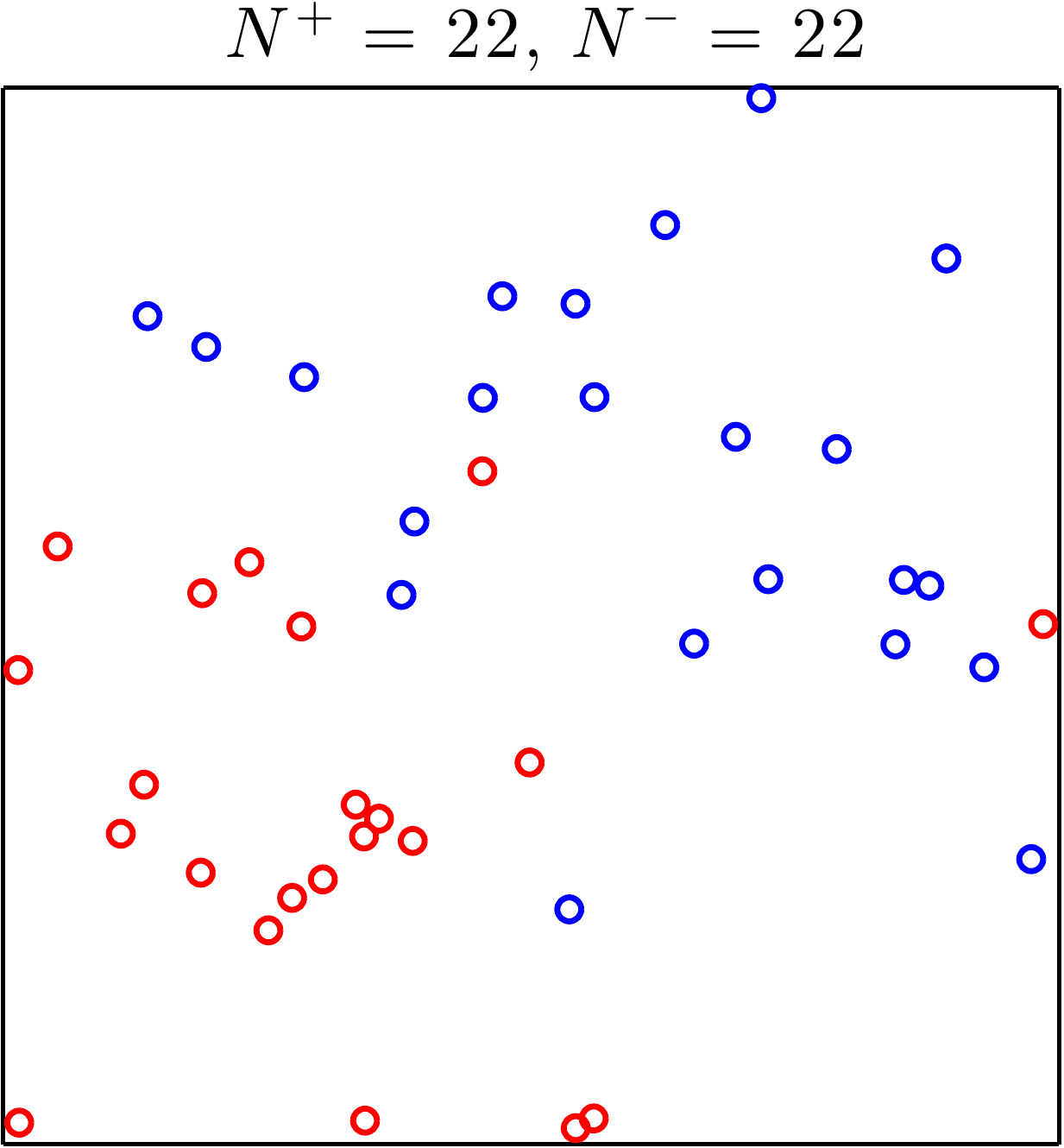}      
      \hspace{0.05cm}
      \includegraphics[width=0.85in]{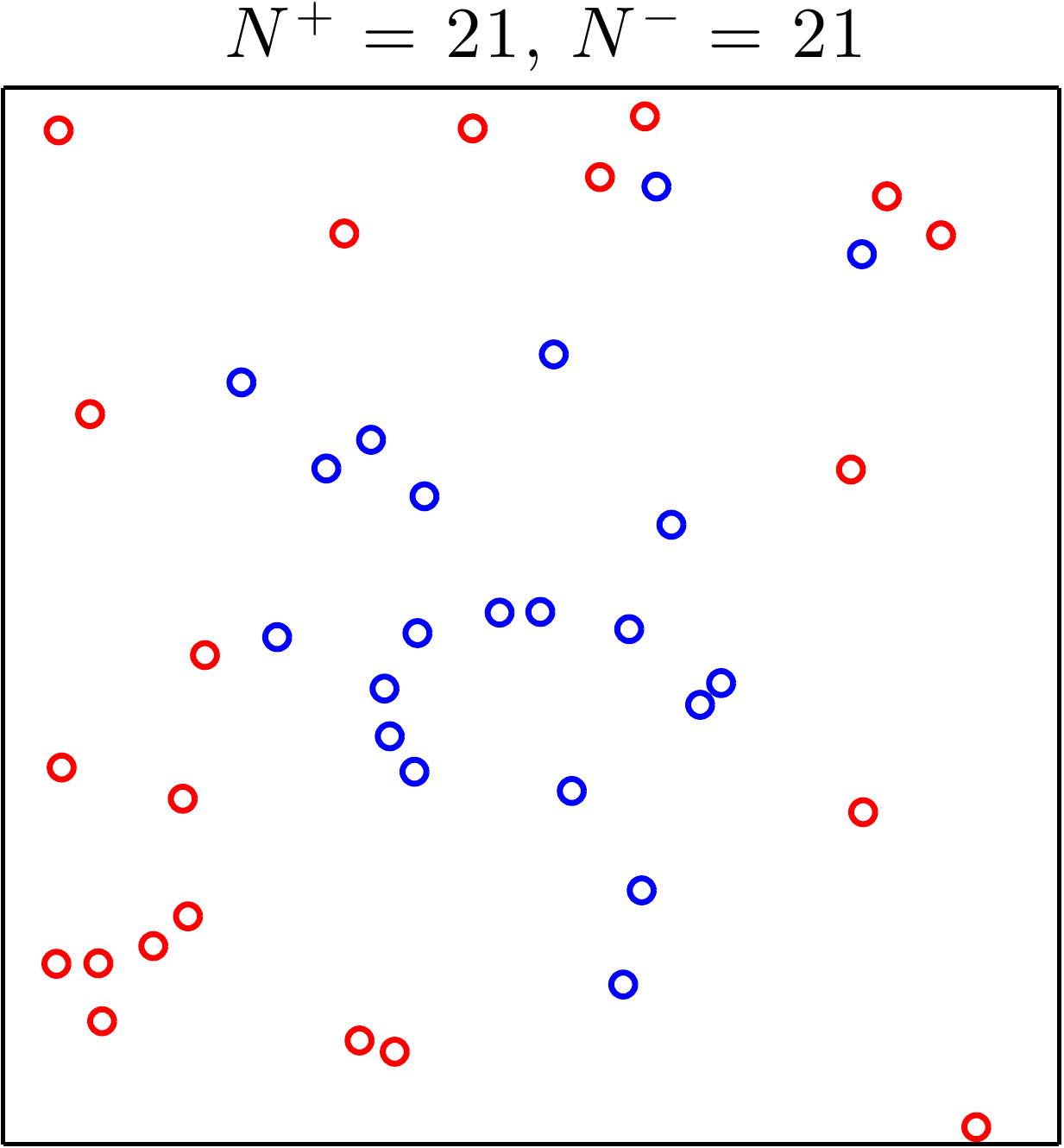}}      
  \end{minipage}
 \begin{minipage}[b]{0.49\textwidth}
    \centering
    \subfigure[\label{fig_t0} \hspace{0.3cm}  $t = 0$, \hspace{1.25cm} $t=1.92t_v$, \hspace{1.35cm} $t = 4.03t_v$ \hfill]{
      \label{fig:mini:subfig:a}
      \includegraphics[width=0.85in]{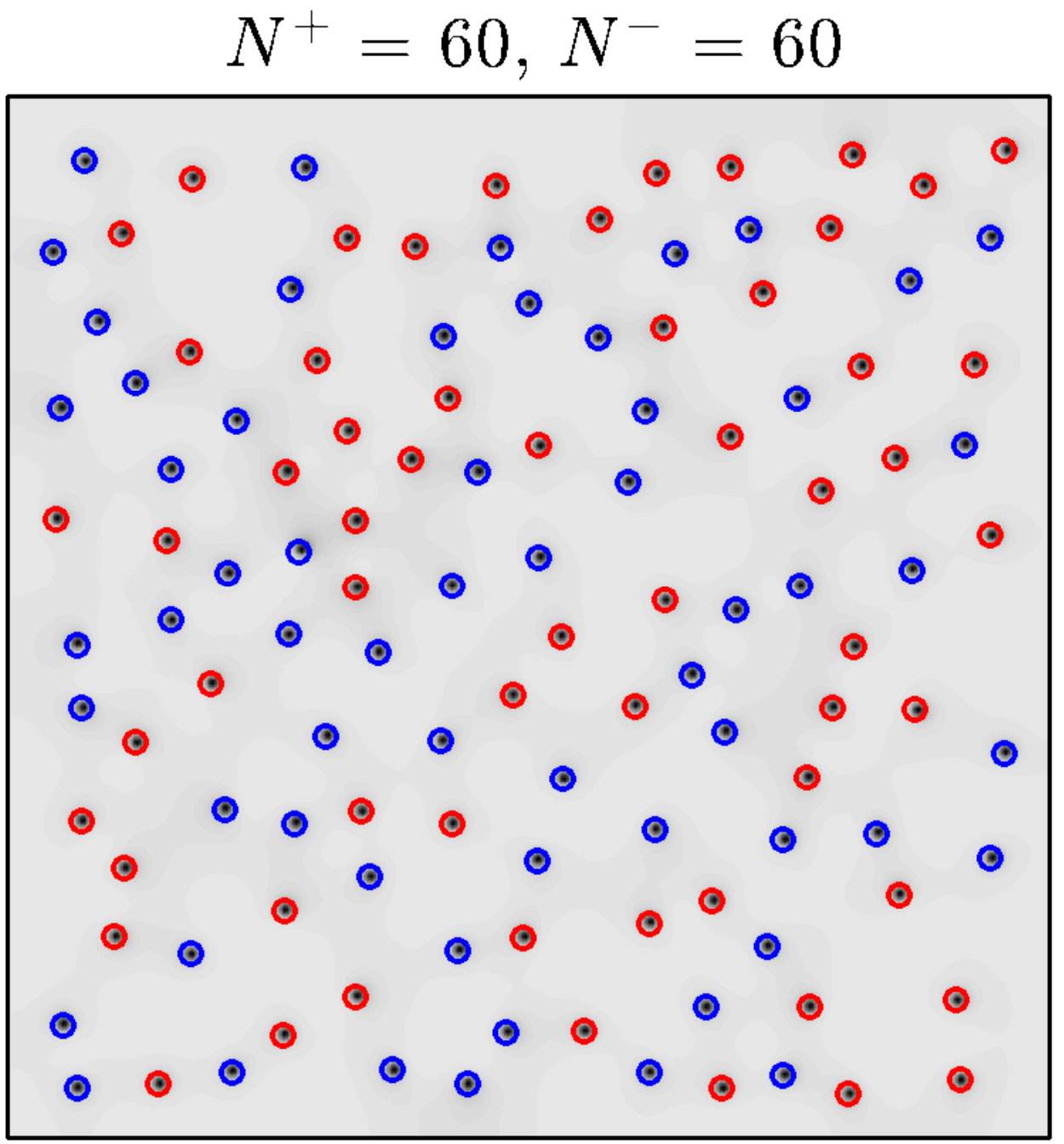}      
      \hspace{0.05cm}
      \includegraphics[width=0.85in]{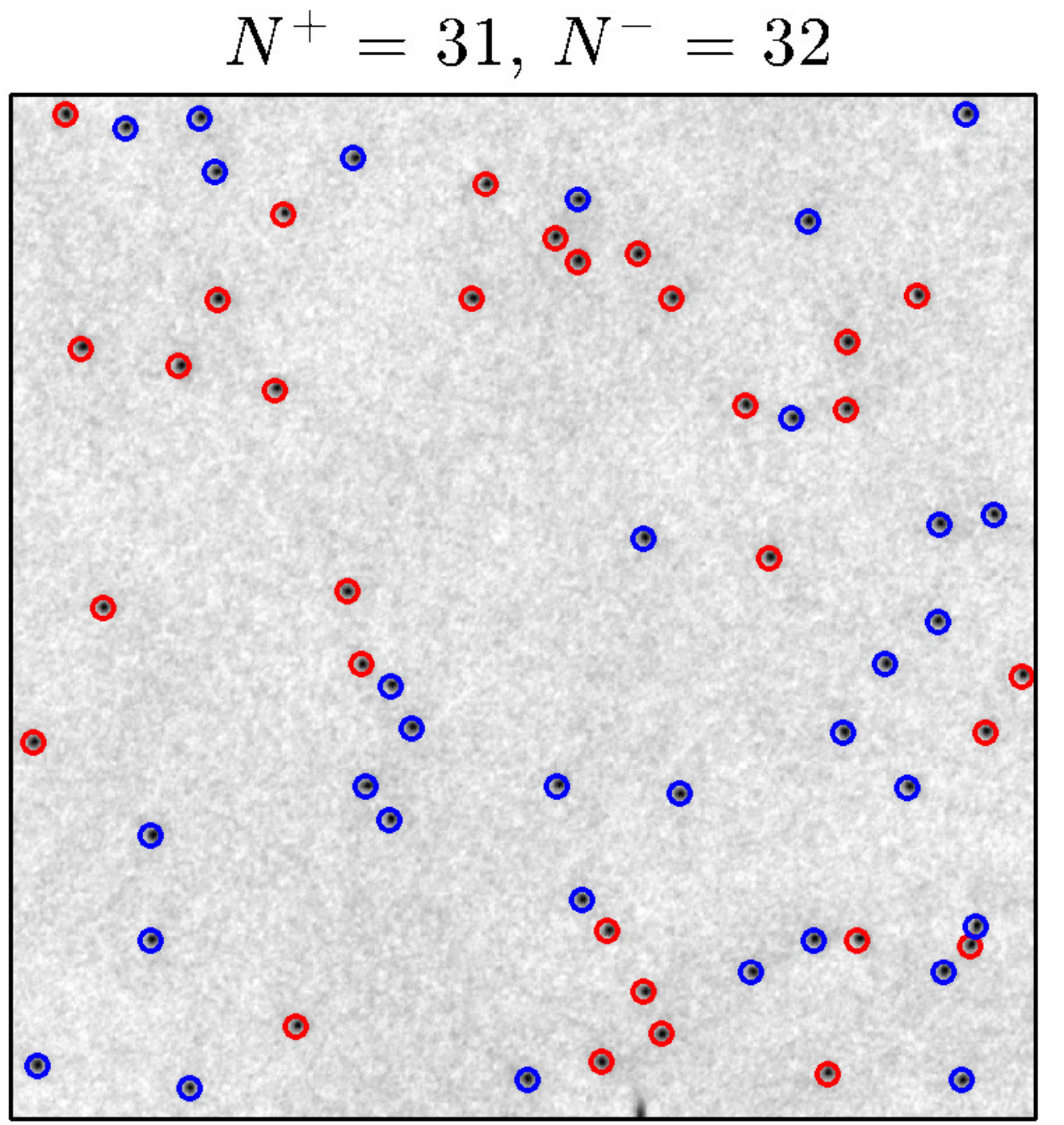}      
      \hspace{0.05cm}
      \includegraphics[width=0.85in]{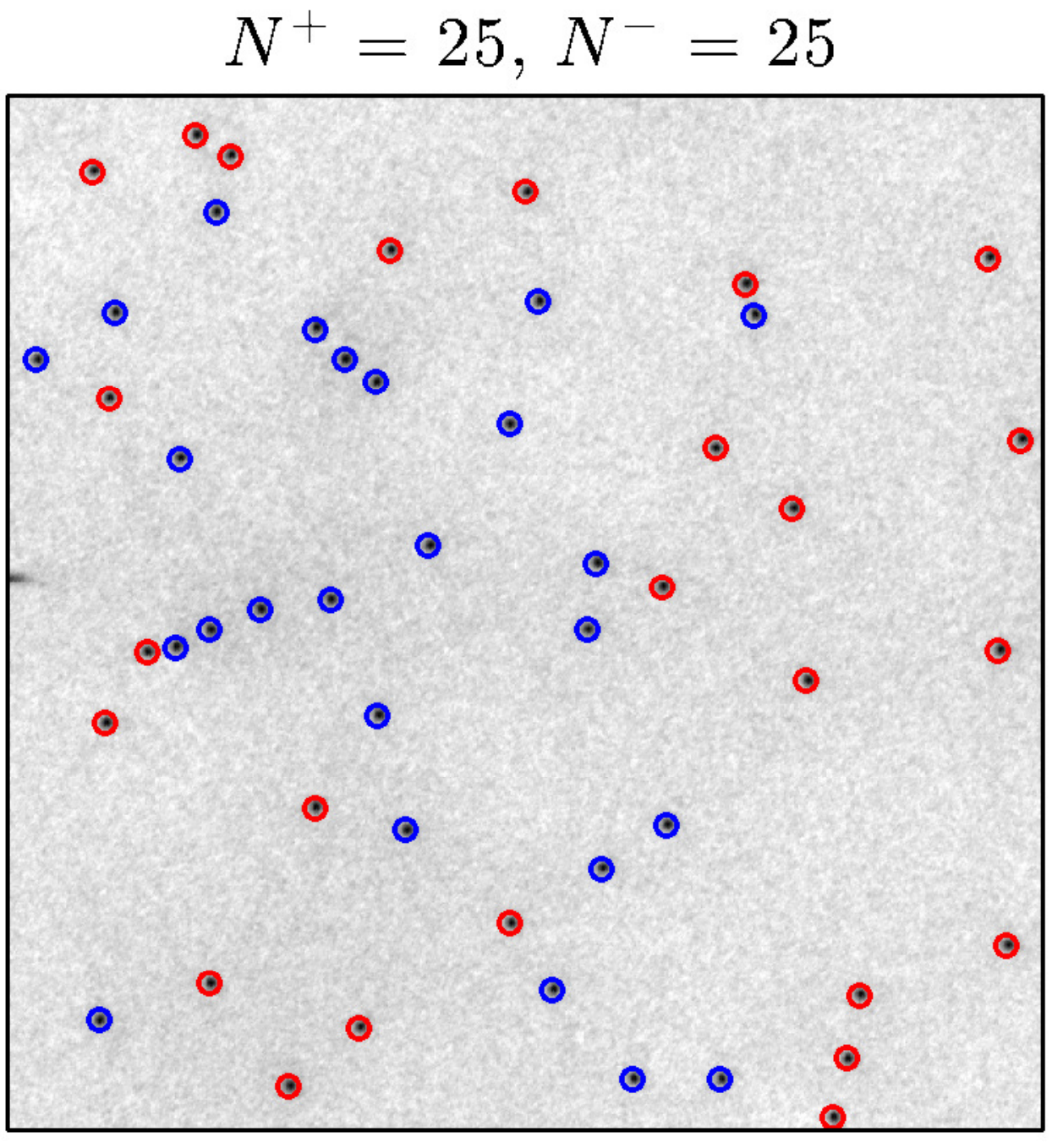}}      
  \end{minipage}  
 \vspace{-0.3cm}
  \caption{Time sequence of locations of vortices (red) and antivortices (blue) for: (a) Point vortex model ($t_v=4$) (b) GP model ($t_v=2.09\times 10^{4}$). In (b), background contour corresponds to plot of $|\phi|^2$.
    \label{fig_PVGPsquare}}
\vspace{-0.4cm}
\end{figure}

Upon integrating our point vortex model forward in time, we obtained a time sequence of the position of point vortices as shown in Fig.\ \ref{fig_PVGPsquare}a. The positions are shown at the initial time, an intermediate time, and a late time in units of $t_v = L^2/|\gamma|$ where $\gamma$ is the non-dimensional circulation of a quantum vortex with winding number one. We note that a monopole distribution emerges at late times in the square domain for our neutral vortex gas. This causes the flow to spontaneously acquire a non-zero value of angular momentum defined as ${\bf L}=\int ( {\bf r} \times \rho {\bf u}) {\rm d}^2{\bf x}$ where ${\bf r}$ is measured relative to the centre of the domain as pointed out in \cite{Taylor2009,Clercx1998,Keetels2008}
(see animations included in \href{run:/Applications/VLC.app/Contents/MacOS/VLC ./supplemental/S1PVDipS.mp4}{S1PVDipS} and \href{run:/Users/Hayder/Downloads/ffmpegXbinaries20060307/mplayer ./supplemental/S1PVMonS.mp4}{S1PVMonS}). These results are consistent with the mean-field predictions which predict that a monopole distribution of vortices is the most probable state in the square and is, therefore, expected to emerge at long times in the latter stages of the simulation.

To further assert the agreement between the theory and the dynamical simulations, we take advantage of the fact that the vortices appear to relax through quasi-equilibrium states and assume ergodicity to replace ensemble averages by time averages. We can then proceed by calculating time-averaged streamfunctions $\Psi = \overline{\psi}$. The instantaneous streamfunction, $\psi$, can be reconstructed from knowledge of vortex/antivortex positions using Eq.\ (\ref{eqn_streamfunction}). 

\begin{figure}[t]
\centering
 \begin{minipage}[b]{0.24\textwidth}
    \centering
    \subfigure[\label{fig_t0} Eigenmode for Point Vortex]{
\includegraphics[width=0.8in]{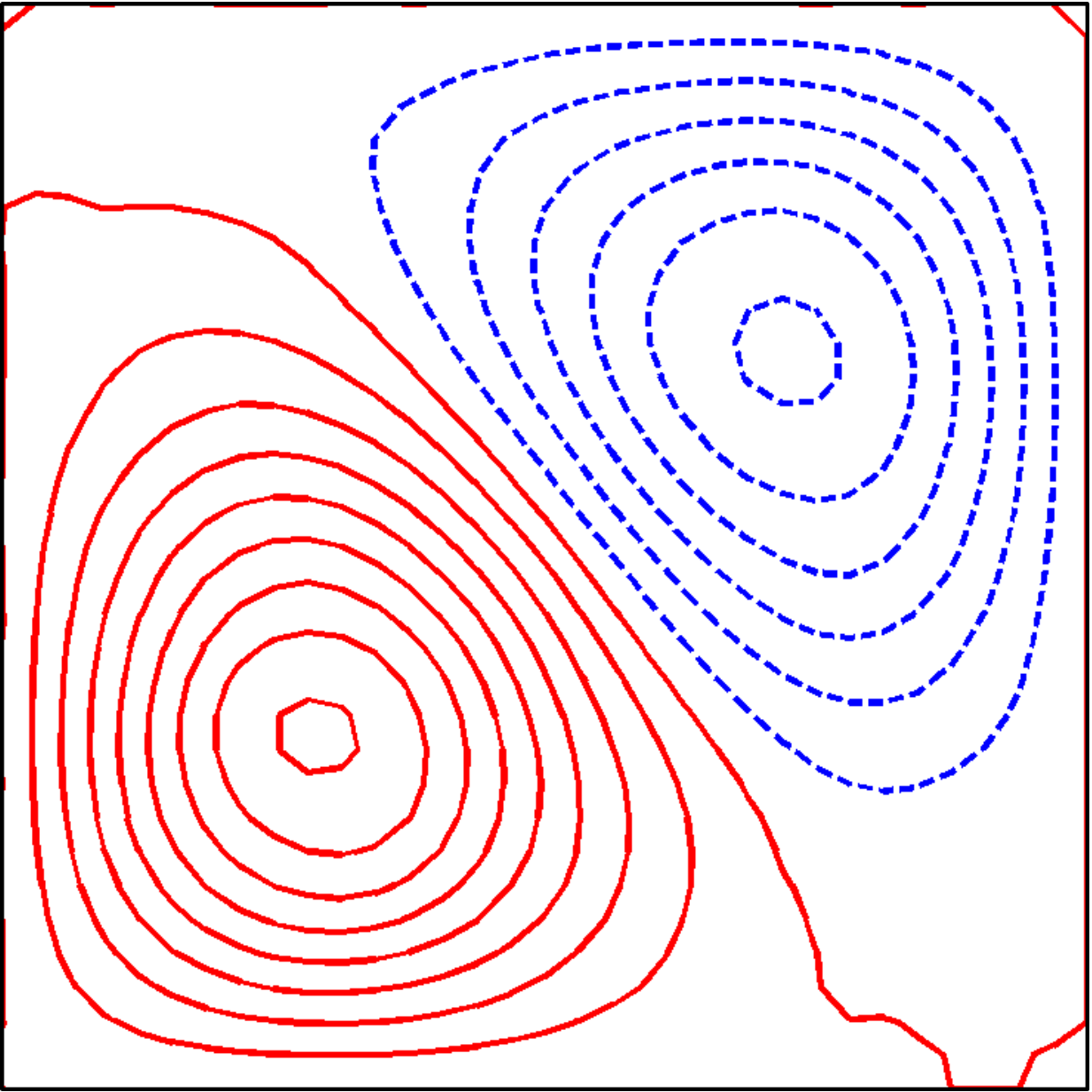}
\includegraphics[width=0.8in]{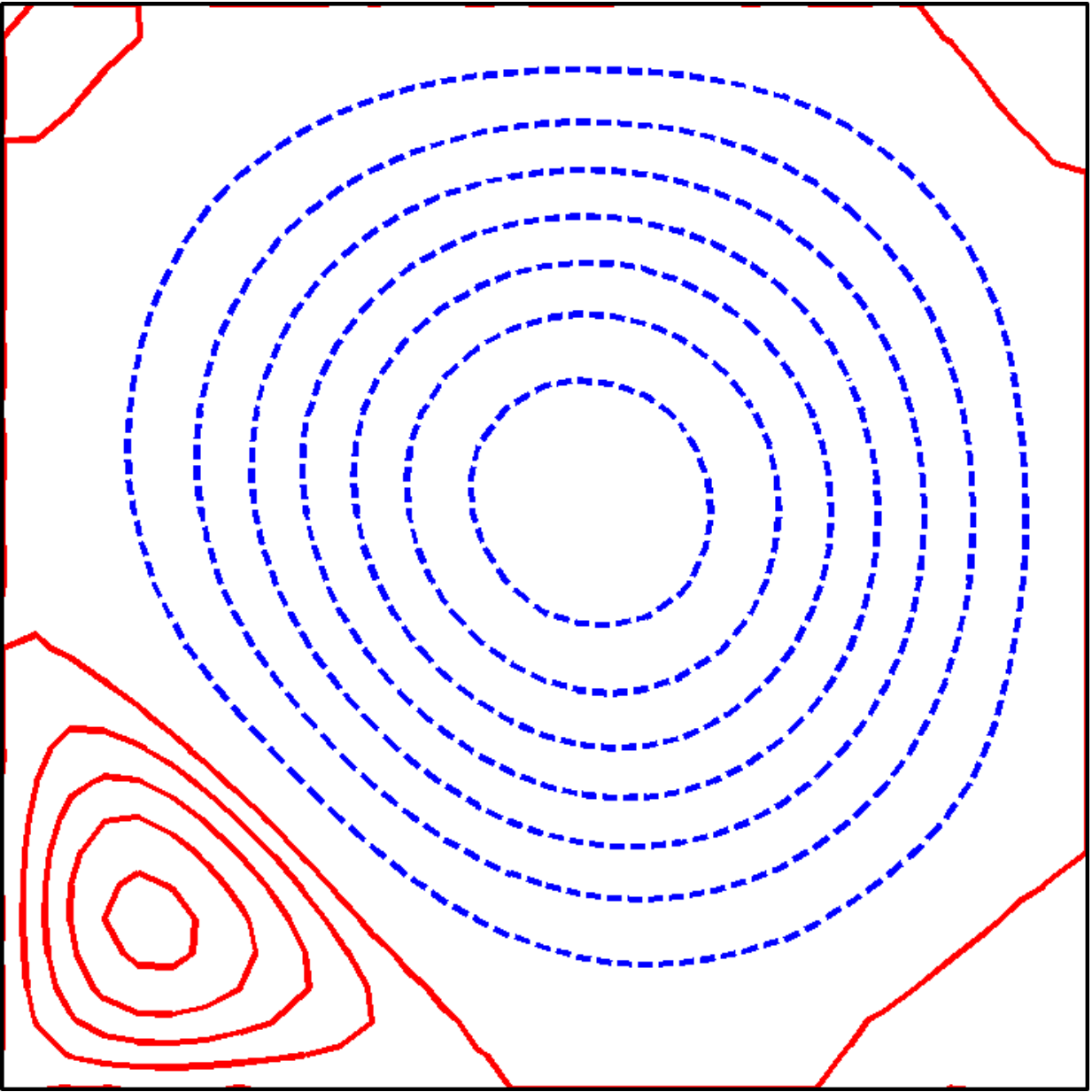}}
  \end{minipage}%
   \begin{minipage}[b]{0.24\textwidth}
    \centering
    \subfigure[\label{fig_t0} Eigenmode for GP]{
\includegraphics[width=0.8in]{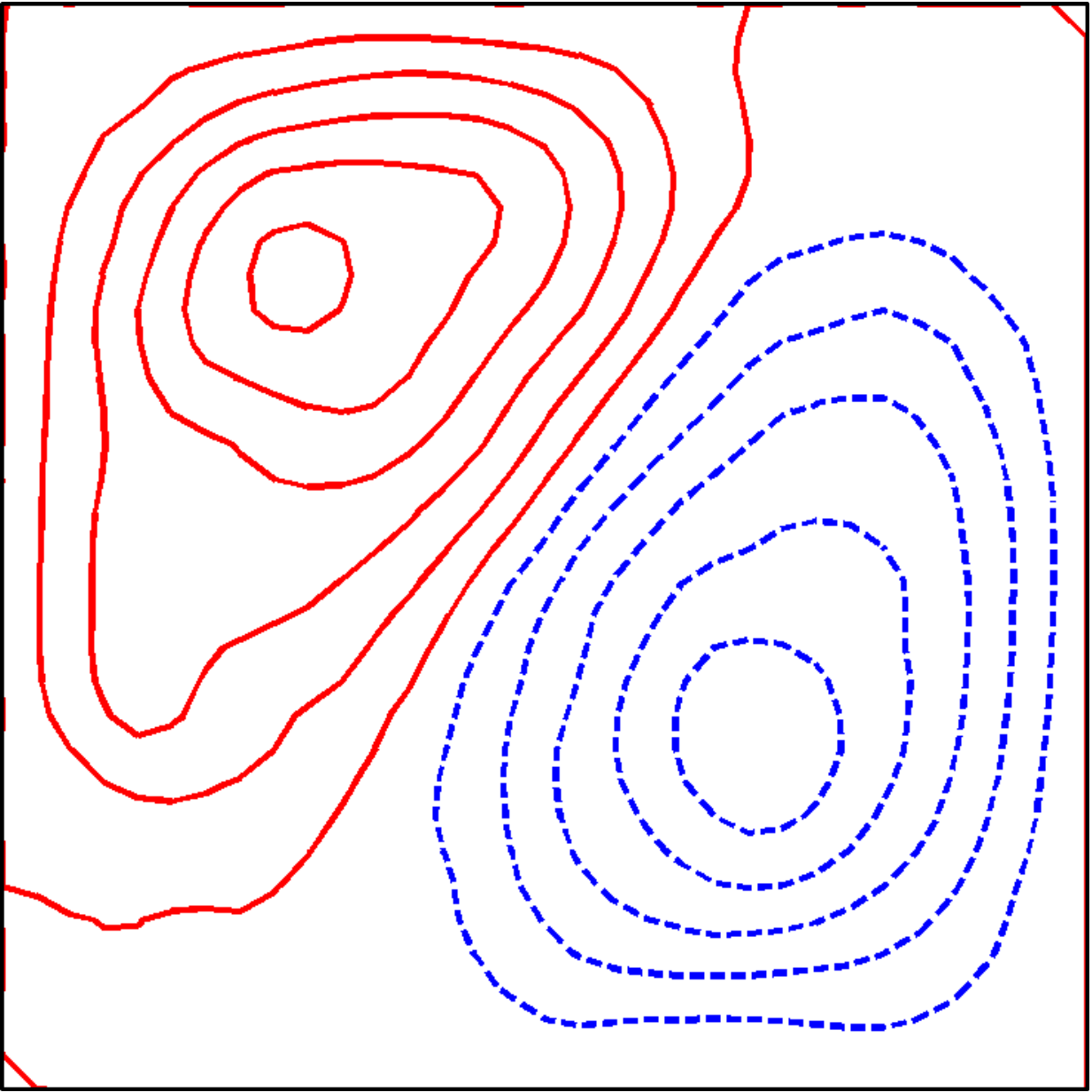}
\includegraphics[width=0.8in]{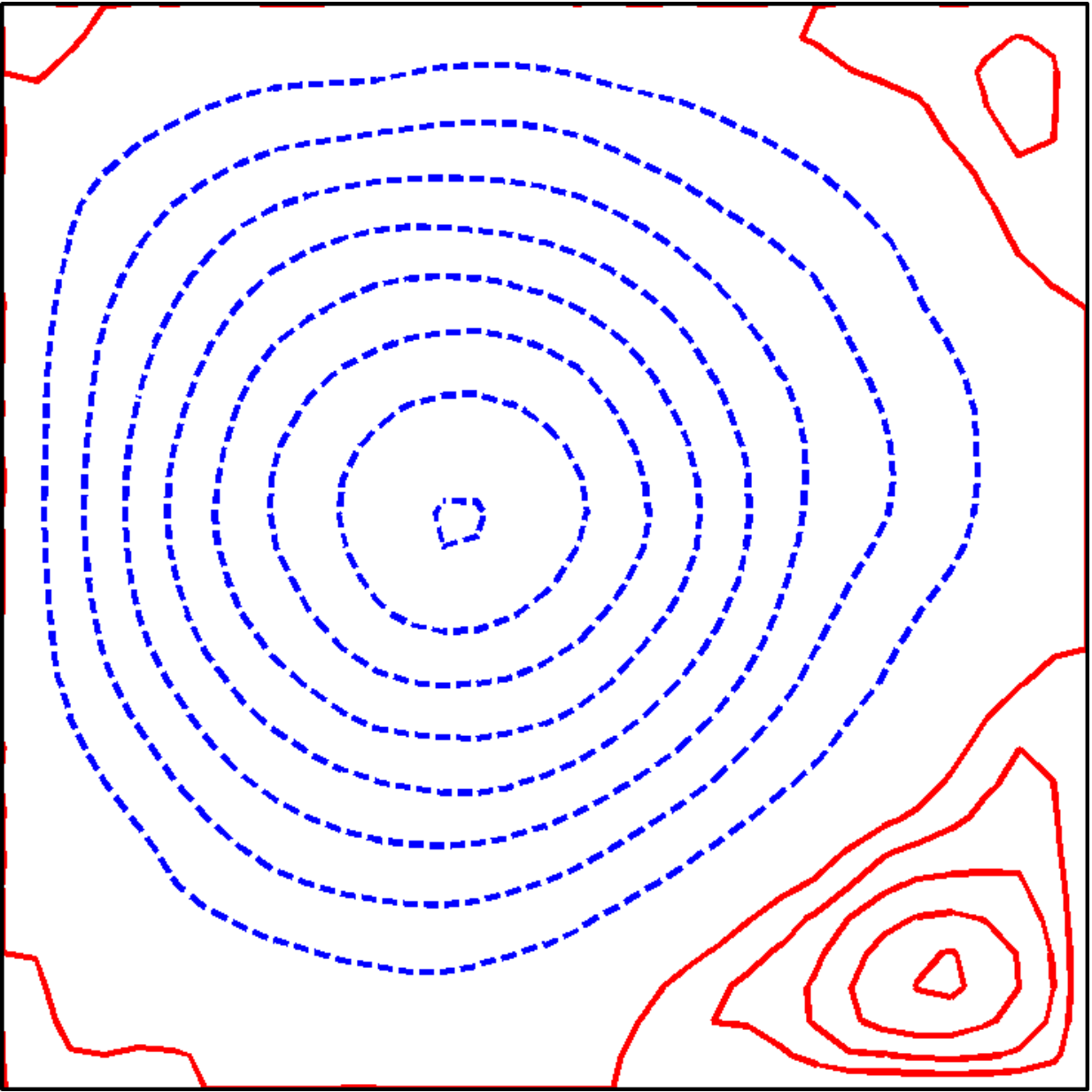}}
  \end{minipage}
  
\centering
 \begin{minipage}[b]{0.24\textwidth}
    \centering
    \subfigure[\label{fig_t0} Streamfunction for Point Vortex]{
\includegraphics[width=0.8in]{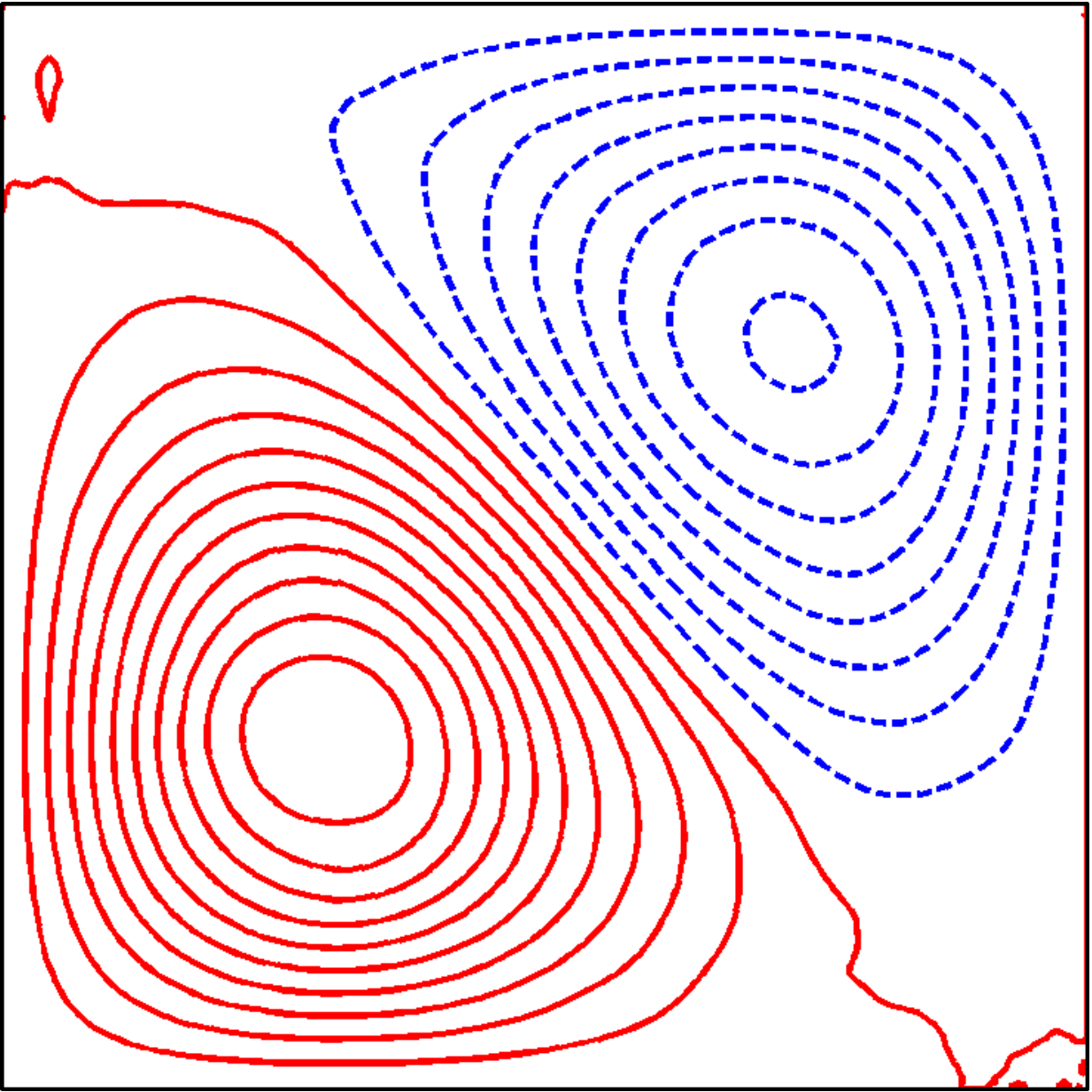}
\includegraphics[width=0.8in]{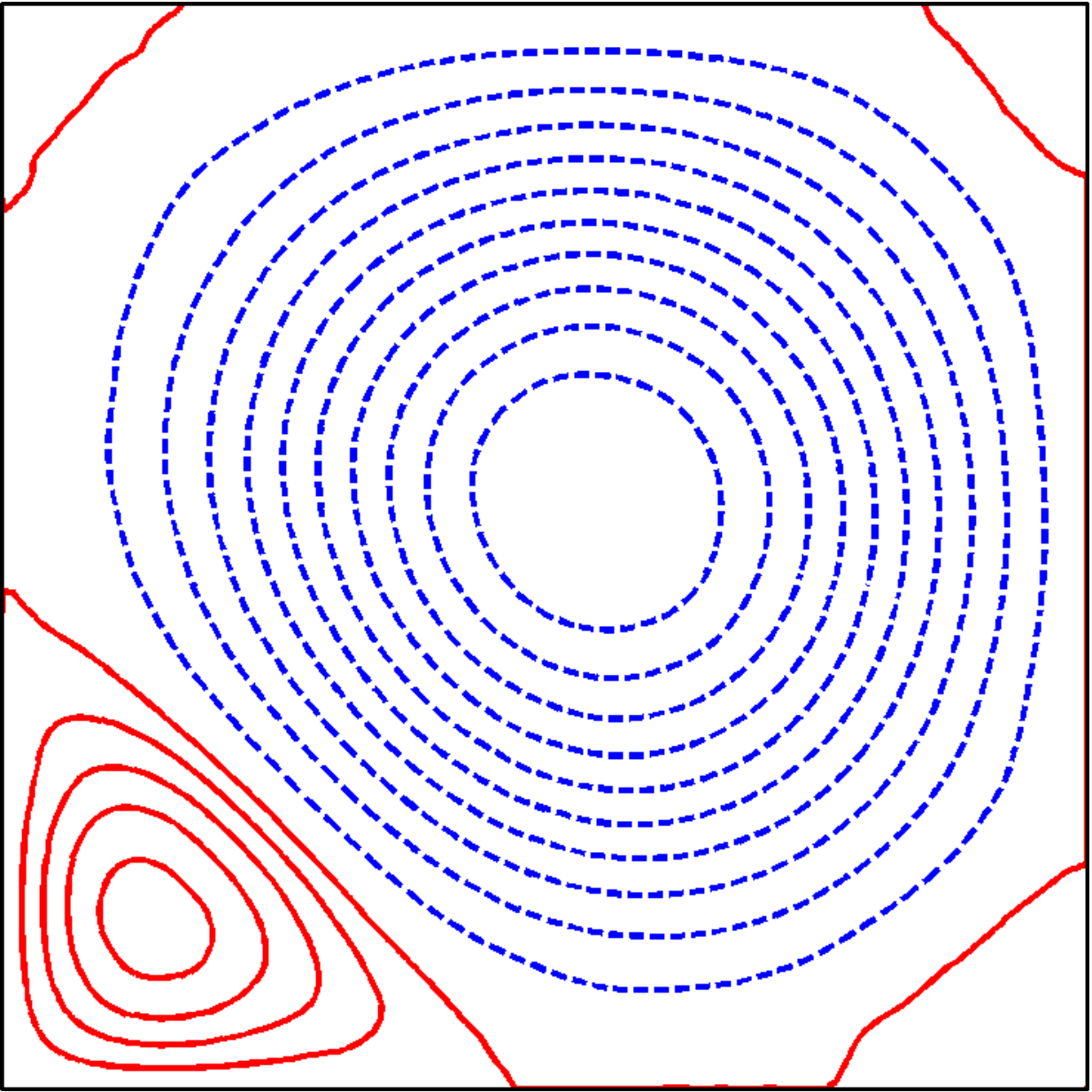}}
  \end{minipage}%
   \begin{minipage}[b]{0.24\textwidth}
    \centering
    \subfigure[\label{fig_t0} Streamfunction for GP]{
\includegraphics[width=0.8in]{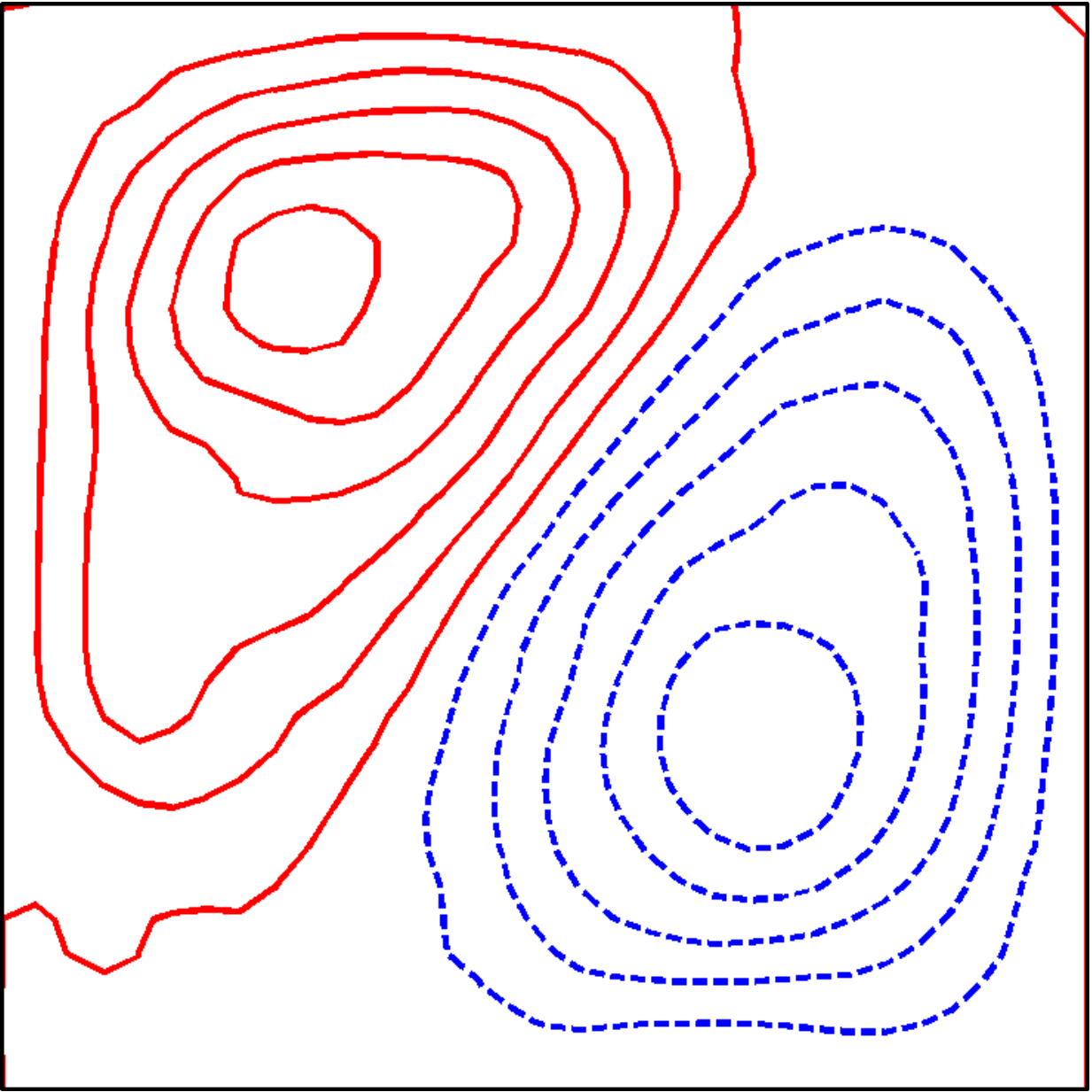}
\includegraphics[width=0.8in]{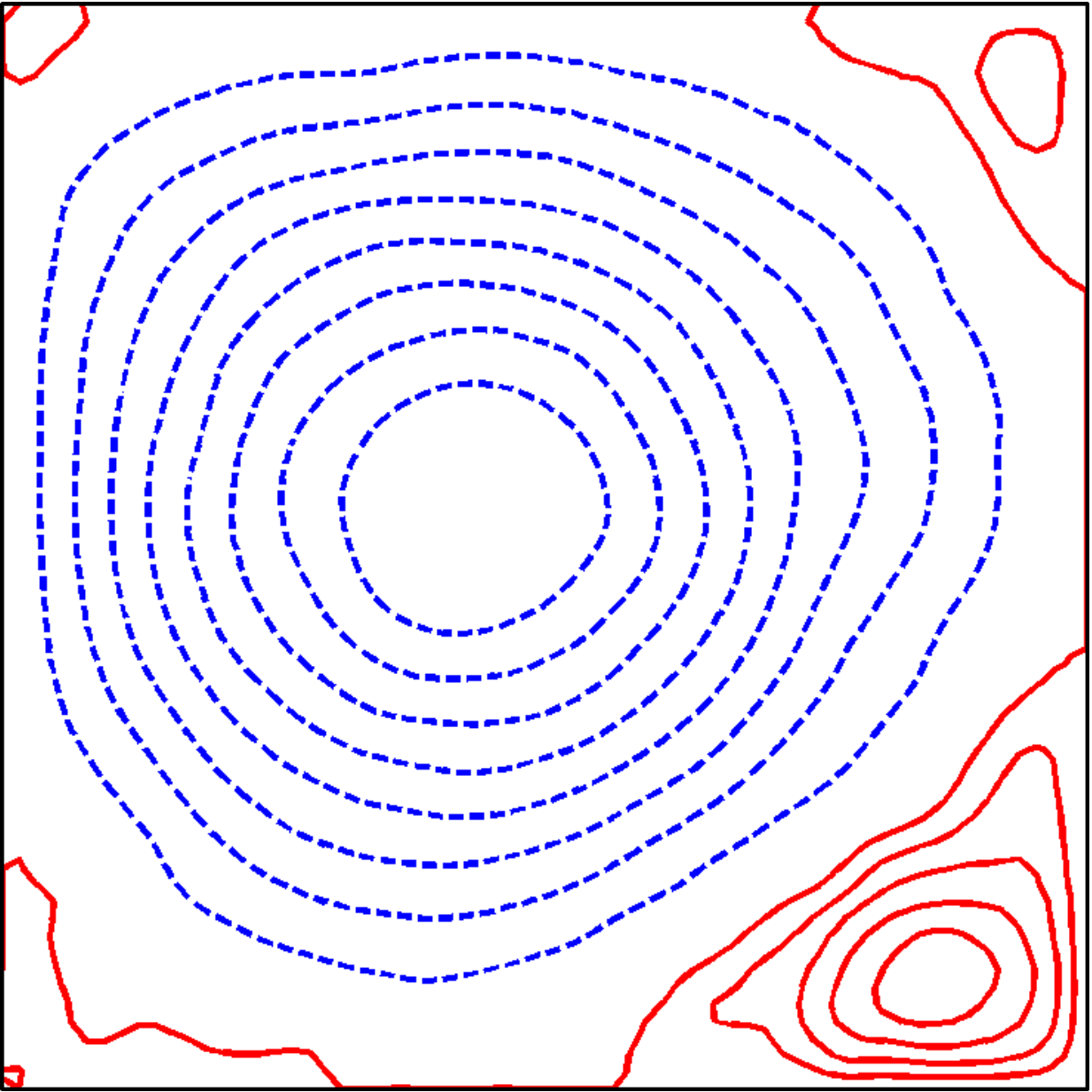}}
  \end{minipage}
 \vspace{-0.3cm}
  \caption{Contour plots in square domain of dominant eigenmode of two point correlator $\big< \psi({\bf x},t) \psi({\bf x}',t)\big>$ for (a) Point vortex model and (b) Gross-Pitaevskii model; Averaged streamlines for (a) Point vortex model; (b) Gross-Pitaevskii model.
In both cases, instantaneous streamfunctions reconstructed from quantized vortex positions.  \label{fig_aveStream_PVGP}}
\end{figure}

In analogy with a BEC \cite{Penrose1956,Goral2002}, a spectral condensate is expected to lead to a non-trivial $\big<{\psi}\big>$ since the condensate can be identified with a maximum eigenvalue of the two-point correlator $\big< \psi({\bf x},t) \psi({\bf x}',t)\big>$ where $\big<  \cdot \big>$, denotes an ensemble average. In this case, the two point correlator can be separated into connected and non-connected parts such that $\big< \psi({\bf x},t) \psi({\bf x}',t)\big>$ = $\Psi({\bf x},t)$ $\Psi({\bf x}',t)$ + $\big< \tilde{\psi}({\bf x},t) \tilde{\psi}({\bf x}',t)\big>$.
In Fig.\ \ref{fig_aveStream_PVGP}a, we present the eigenmode corresponding to the maximum eigenvalue of the two-point correlator. The corresponding time-averaged mean of the wavefunction is presented in Fig.\ \ref{fig_aveStream_PVGP}c. The expectations are evaluated over the time intervals, 
$t \in [45.87-48.07,64.40-67.54]$, which coincide with the animations (see \href{run:/Applications/VLC.app/Contents/MacOS/VLC ./supplemental/S1PVDipS.mp4}{S1PVDipS} and \href{run:/Users/Hayder/Downloads/ffmpegXbinaries20060307/mplayer ./supplemental/S1PVMonS.mp4}{S1PVMonS}). 
We observe that the dominant eigenmodes extracted from an eigenvalue decomposition of the two point correlator over the two time intervals are similar to the streamfunctions that are obtained from the time-averaging. Moreover, both fields coincide with the predictions of the mean-field theory. When interpreted together, these results provide clear evidence that our definition of the spectral condensate, that is analogous to the Penrose-Onsager definition of a Bose-Einstein condensate, allows us to clearly identify the 
the spontaneous emergence of a coherent mean flow in our simulations. Moreover, it confirms that a dipole emerges at intermediate times which gives way to the monopole at later times.

We have simulated the same initial condition with the GP model given in Eq.\ (\ref{eqn_GP}) using the parameters given above. The resulting initial density field is shown in Fig.\ \ref{fig_PVGPsquare}b together with the vortex positions indicated in red for vortices and blue for antivortices. 
We observe that there is a clear disparity between the relaxation time-scales of the point vortex and the GP models even though we have scaled the results of the two simulations with the vortex time-scales.
We note that, in the GP model, vortices can interact with phonons and other wave excitations that are absent in the point vortex simulations. These are produced within the system during the motion of the vortices that tend to radiate energy or during the annihilation of vortex-antivortex pairs, a mechanism that abruptly converts incompressible energy to compressible energy. The effect of these waves that continue to reside within an isolated system is to modify the vortex dynamics by introducing dissipative effects on their motion. Consequently the vortex gas in the GP model will relax by evolving through lower energy states (of the vortex system) in comparison to the point vortex model with a consequent effect on the relaxation time scale.

As with the point vortex model, the streamfunction can be reconstructed from knowledge of the location of vortices and antivortices that we have from the simulations of the GP model. The dominant eigenmode of the two point correlator and time-averages of the streamfunctions within the intervals $t \in [1.92t_v-2.68t_v,3.64t_v-5.56t_v]$ are presented in Fig.\ \ref{fig_aveStream_PVGP}b and \ref{fig_aveStream_PVGP}d, respectively. As with the point vortex simulations, we observe clear evidence of the dipole at intermediate times followed by a monopole distribution of vortices at longer times in agreement with the mean-field theory (see \href{run:/Applications/VLC.app/Contents/MacOS/VLC ./supplemental/S1GPDipS.mp4}{S1GPDipS}
and \href{run:/Users/Hayder/Downloads/ffmpegXbinaries20060307/mplayer ./supplemental/S1GPMonS.mp4}{S1GPMonS}). 
We note that the coherent flow patterns that emerge during the relaxation of our system bear remarkable similarity with the large scale flows observed in forced classical fluid experiments due to the spectral condensation of energy (see Fig.\ 1c of \cite{Xia2011,Xia2009}).

\begin{figure}[t]
\centering
 \begin{minipage}[b]{0.49\textwidth}
    \centering
    \subfigure[\label{fig_t0} \hspace{0.3cm}  $t = 0$, \hspace{1.25cm} $t=38.04t_v$, \hspace{1.35cm} $t = 77.6t_v$ \hfill]{
      \label{fig:mini:subfig:a}
      \includegraphics[width=1.05in]{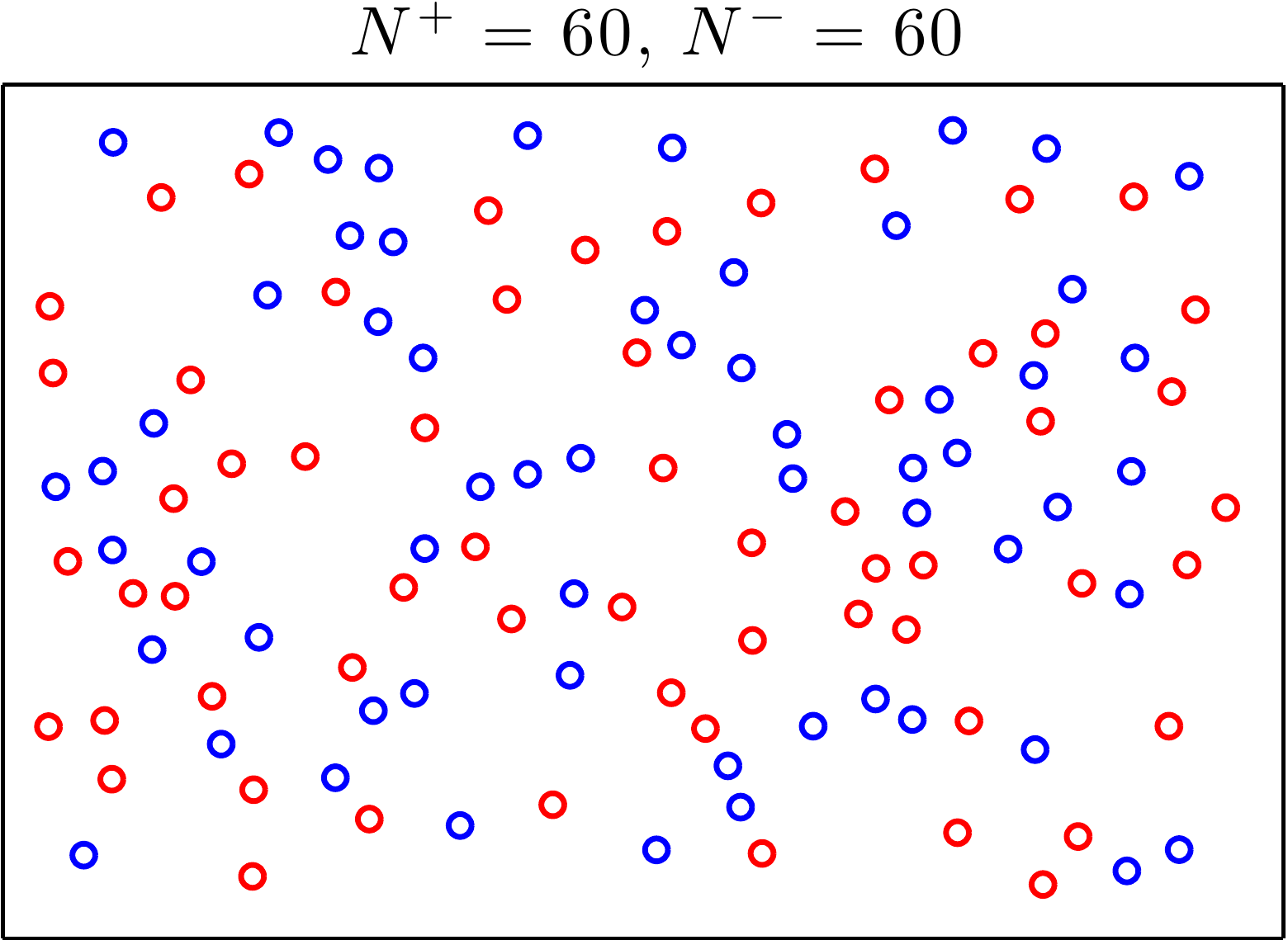}      
      \includegraphics[width=1.05in]{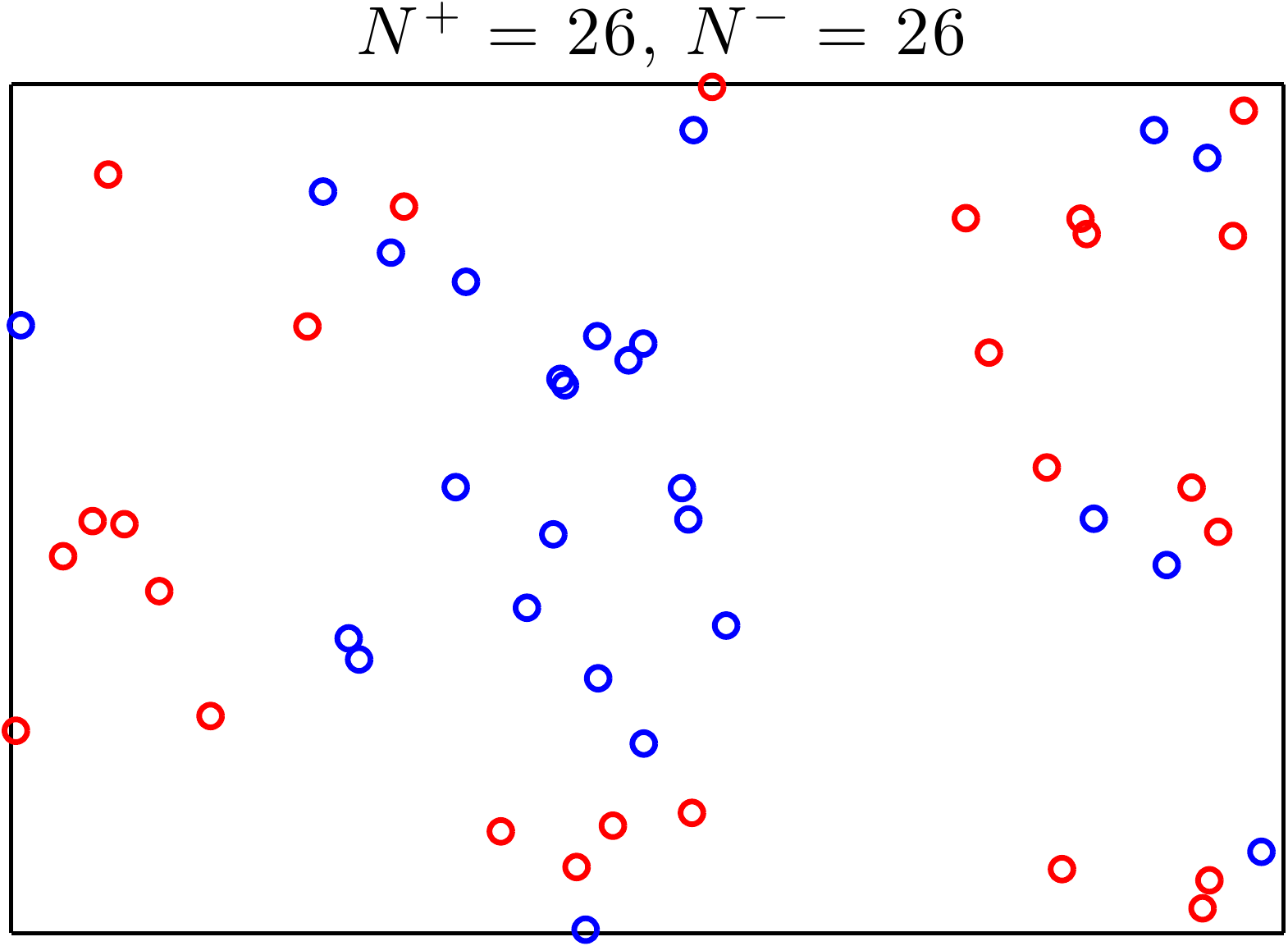}      
      \hspace{0.05cm}
      \includegraphics[width=1.05in]{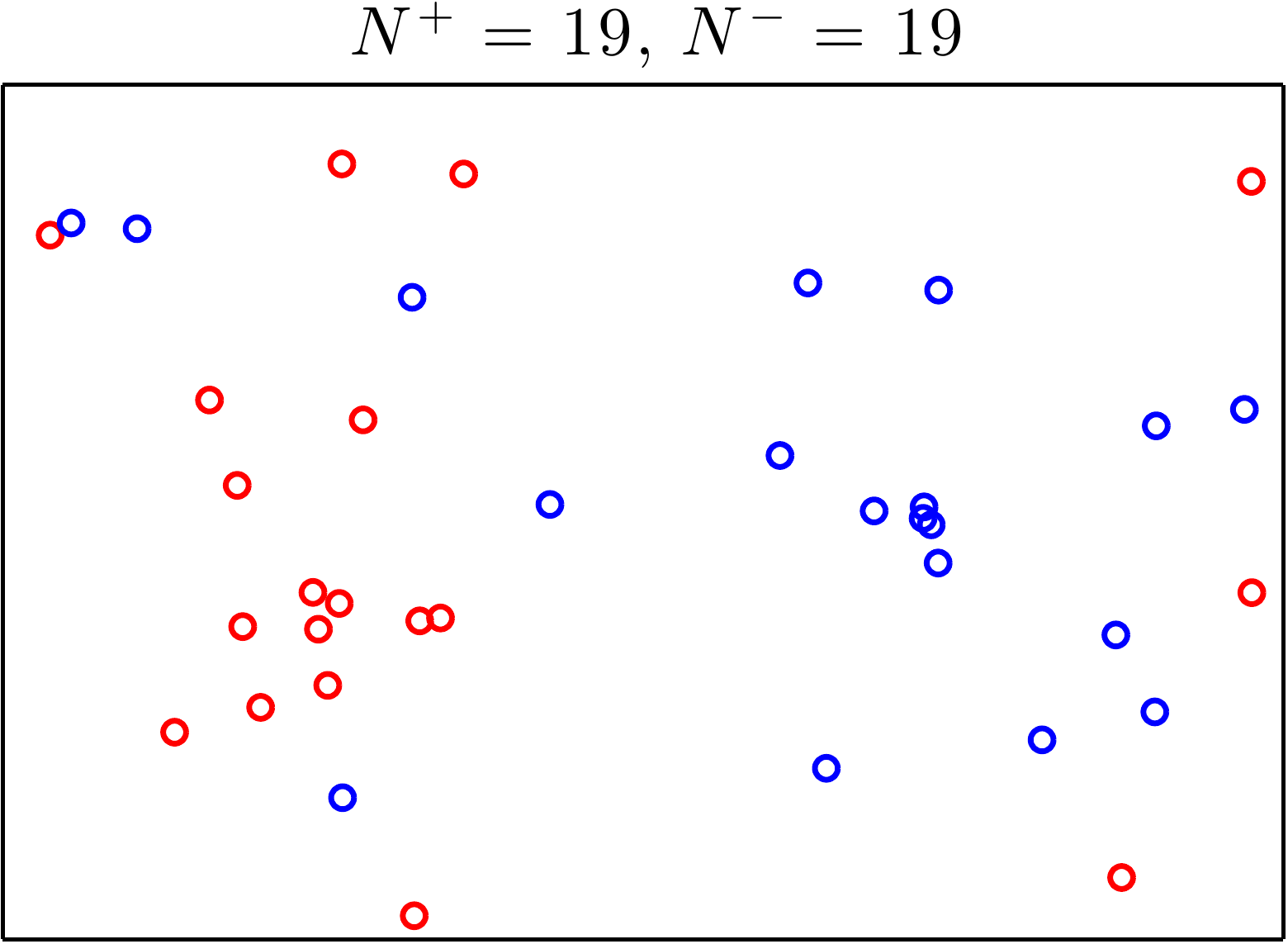}}      
      \hspace{0.05cm}   
  \end{minipage}
 \begin{minipage}[b]{0.49\textwidth}
    \centering
    \subfigure[\label{fig_t0} \hspace{0.3cm}  $t = 0$, \hspace{1.25cm} $t=2.88t_v$, \hspace{1.35cm} $t = 5.75t_v$ \hfill]{
      \label{fig:mini:subfig:a}
      \includegraphics[width=1.05in]{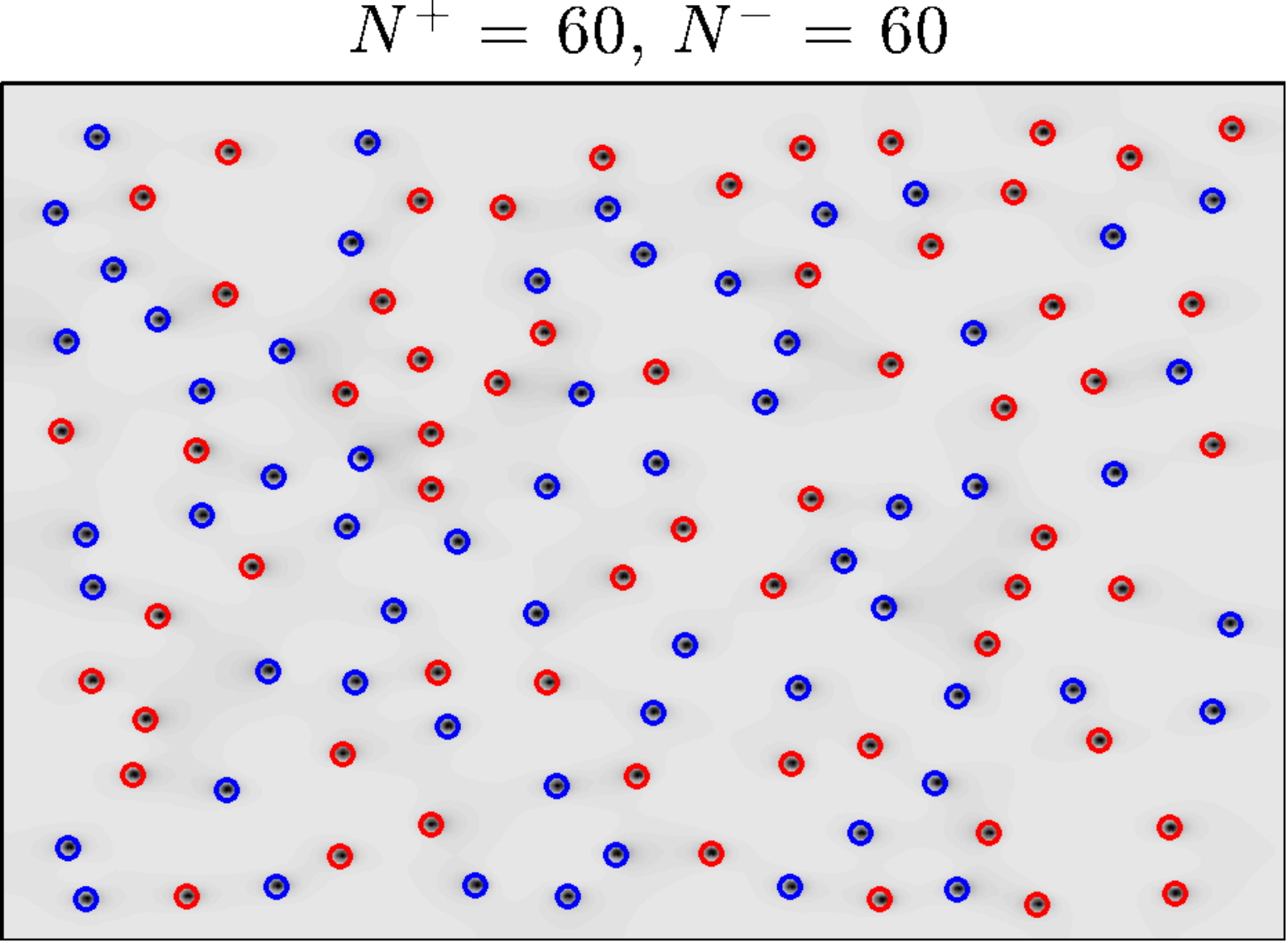}      
      \includegraphics[width=1.05in]{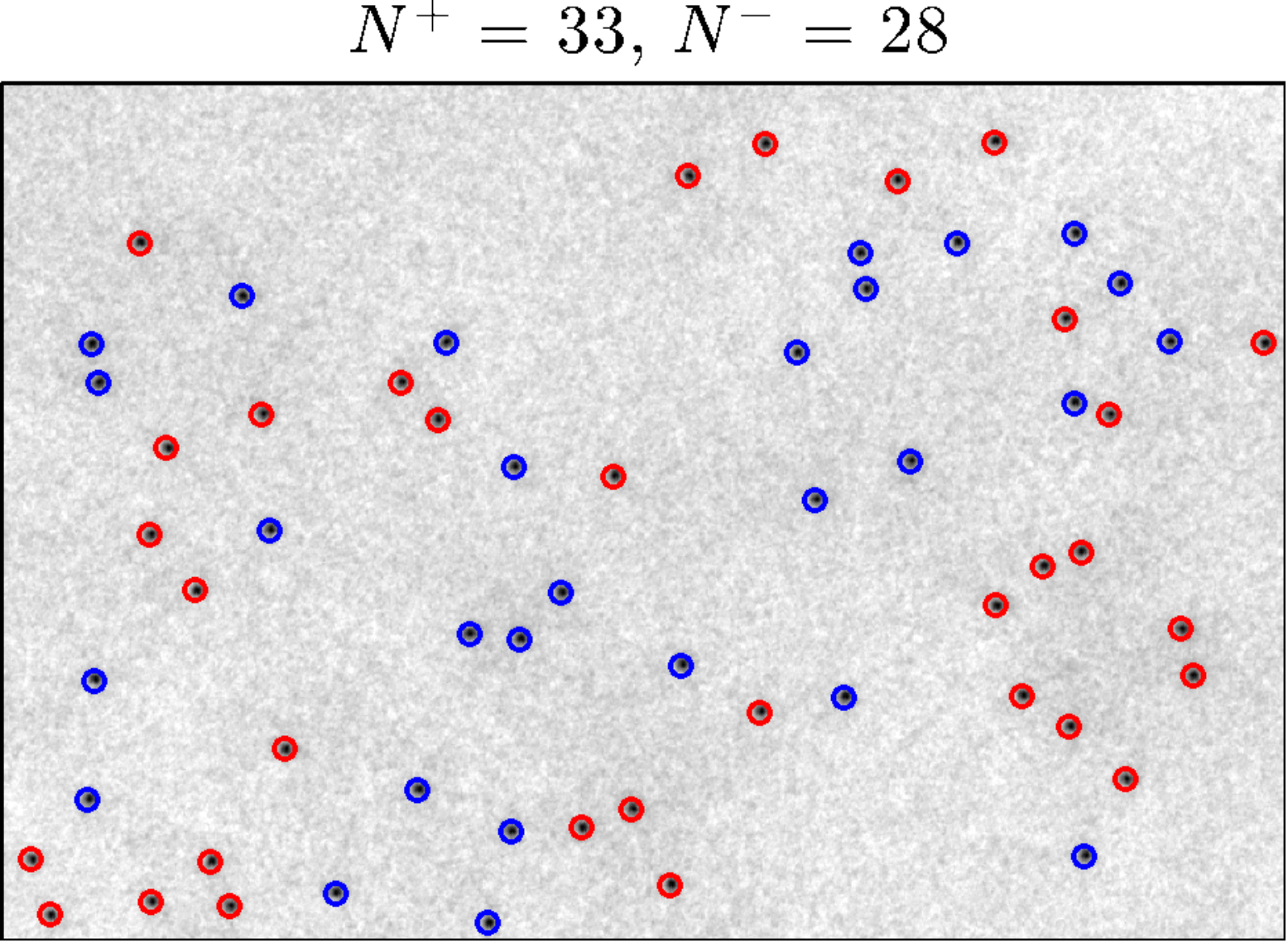}      
      \hspace{0.05cm}
      \includegraphics[width=1.05in]{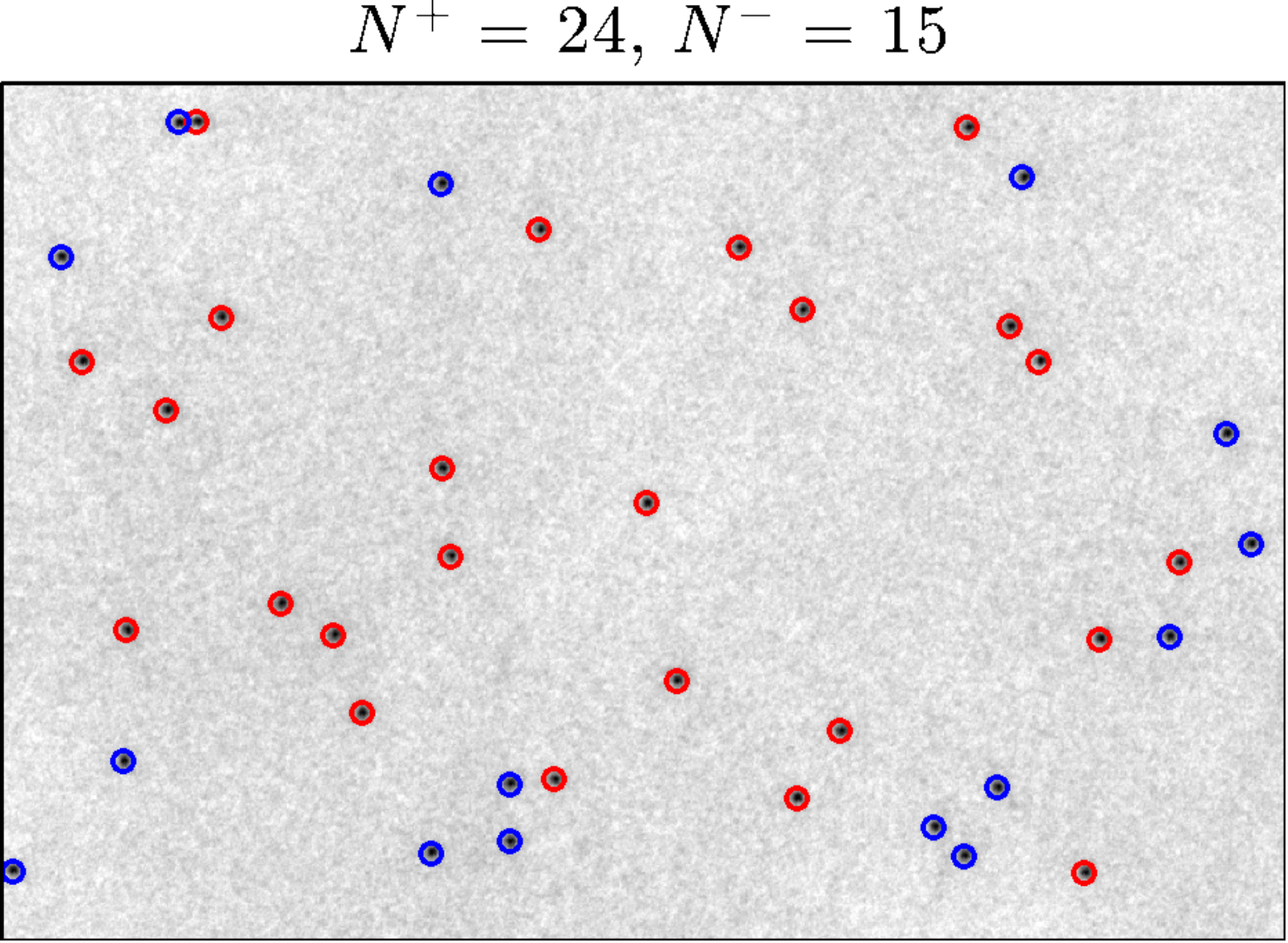}}      
      \hspace{0.05cm}   
  \end{minipage}  
 \vspace{-0.3cm}
  \caption{Time sequence of locations of vortices (red) and antivortices (blue) for: (a) Point vortex model ($t_v=4$) (b) GP model ($t_v=2.09\times 10^{4}$).
  In (b), background contour corresponds to plot of $|\phi|^2$.
    \label{fig_PVGPrectangle}}
\vspace{-0.4cm}
\end{figure}

Because of the long-range Coulomb like interactions, our mean-field predictions presented above for the rectangle indicate that a dipole is expected to emerge for a rectangular flow.
We have checked these mean-field predictions by simulating the dynamics of vortices using both the point vortex and the GP models for a rectangular domain with an aspect ratio equal to $1.5$ while keeping the area fixed to that of the square. 

\begin{figure}[t]
\centering
 \begin{minipage}[b]{0.49\textwidth}
    \centering
    \subfigure[\label{fig_t0} Point Vortex]{
\includegraphics[width=1.05in]{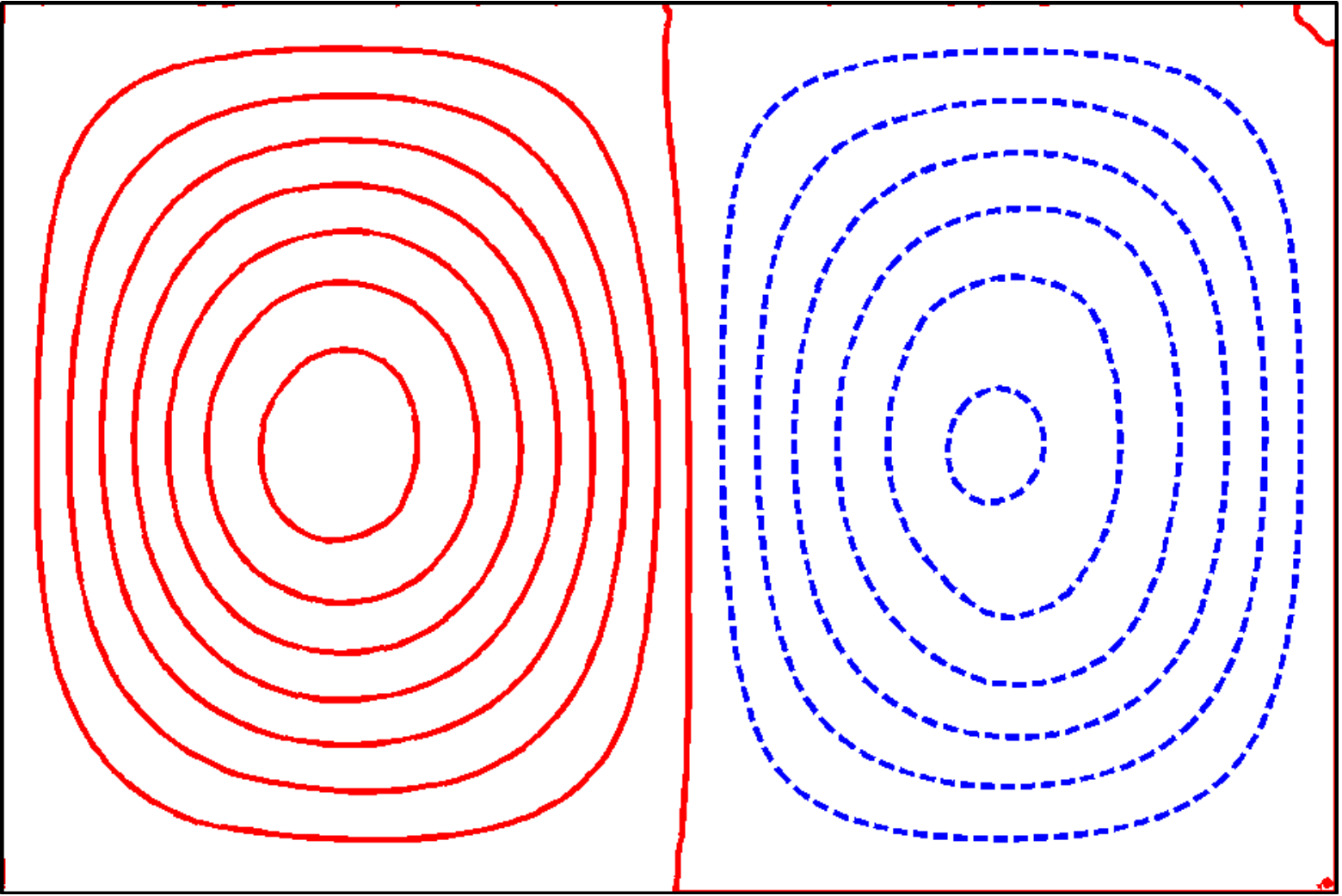}}
    \subfigure[\label{fig_t0} GP (Case1)]{
\includegraphics[width=1.05in]{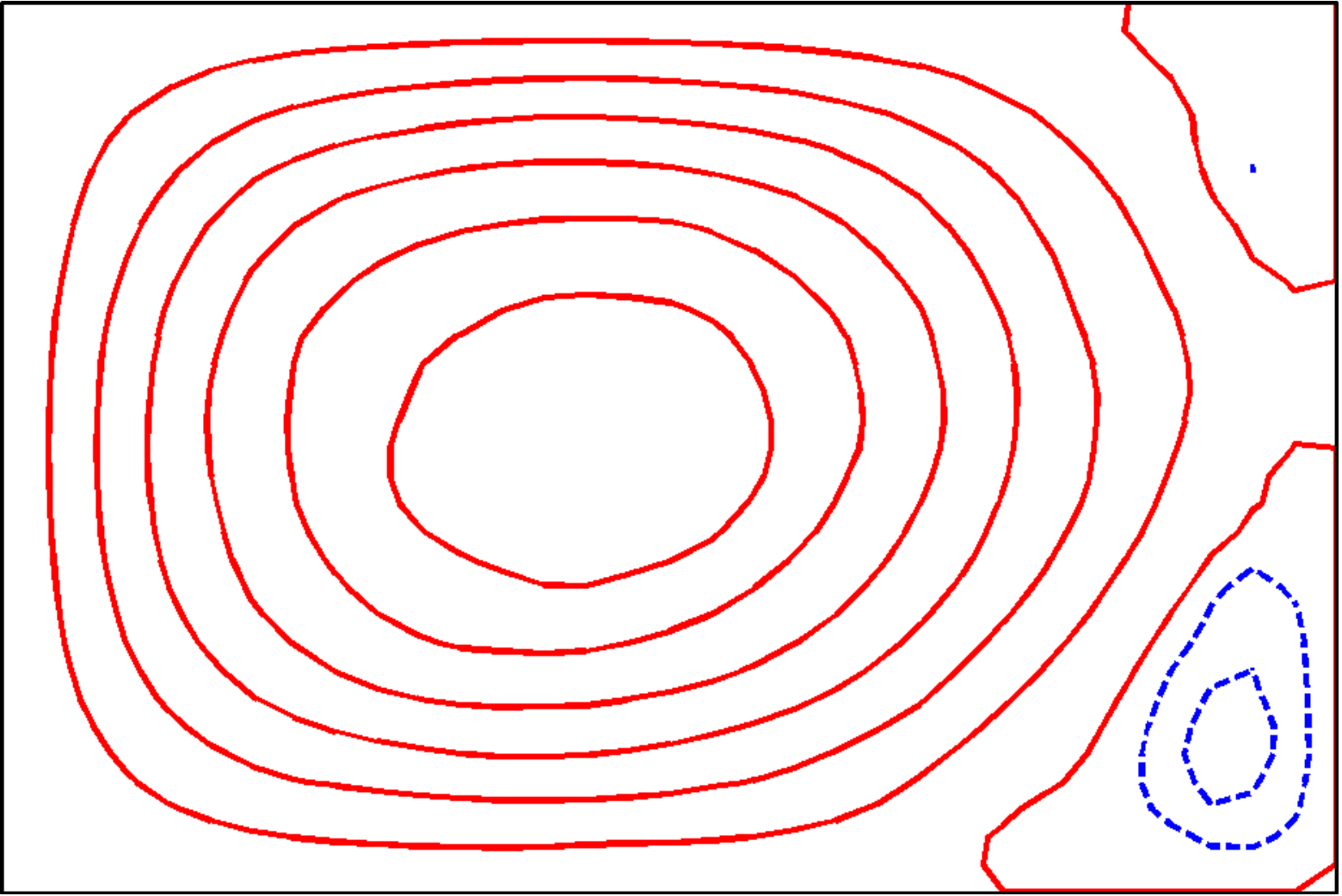}}
    \subfigure[\label{fig_t0} GP (Case 2)]{
\includegraphics[width=1.05in]{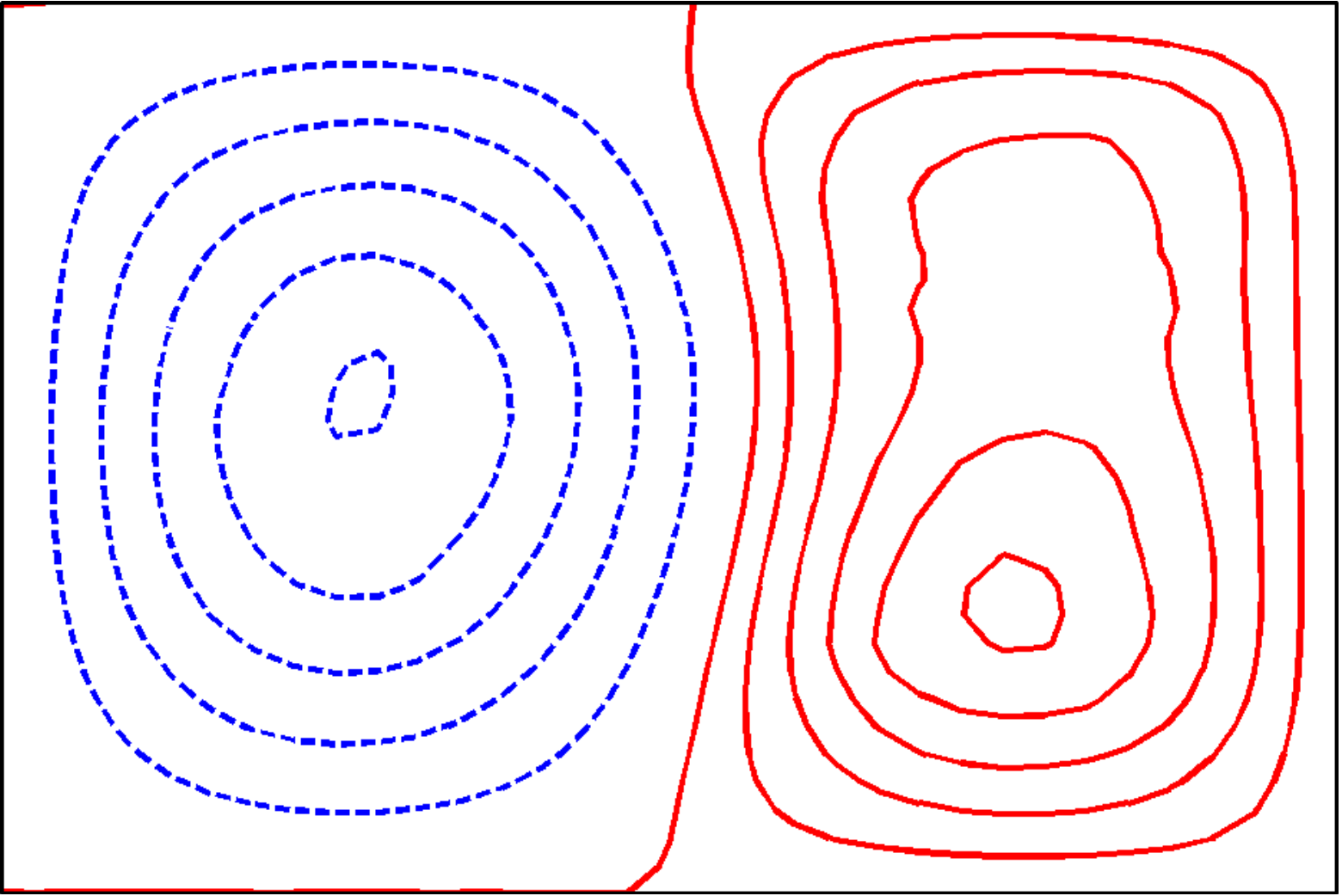}}
  \end{minipage}
 \vspace{-0.3cm}
  \caption{Averaged streaminlines in rectangular domain calculated from quantized vortex positions for: (a) point vortex model; (b) and (c) GP simulations with different parameters. \label{fig_aveStream_PVGPrectangle}}
\vspace{-0.4cm}
\end{figure}

Time sequence plots showing the positions of the vortices in both models for the rectangle are presented in Fig.\ \ref{fig_PVGPrectangle}. In contrast to the square, the long time behaviour leads to a dipole distribution in the point vortex model. By comparison, for the GP simulations, the monopole appears to persist throughout the simulation with no evidence of the dipole emerging. (We refer to \href{run:/Applications/VLC.app/Contents/MacOS/VLC ./supplemental/S2PVDipR.mp4}{S2PVDipR} and \href{run:/Users/Hayder/Downloads/ffmpegXbinaries20060307/mplayer ./supplemental/S2GPMonR.mp4}{S2GPMonR} for a further illustration). This is further confirmed from the time-averaged streamfunctions presented in Figs.\ \ref{fig_aveStream_PVGPrectangle}a, and b which were evaluated over the time intervals 
$t\in [75.4t_v-78.54t_v]$ and $t \in [5.75t_v-6.71t_v]$ for the point vortex and GP models, respectively. 

\begin{figure}[t]
\centering
\hspace{-0.05cm}
 \begin{minipage}[b]{0.49\textwidth}
    \centering
    \subfigure[\label{fig_t0} Square]{
\includegraphics[width=2.68in]{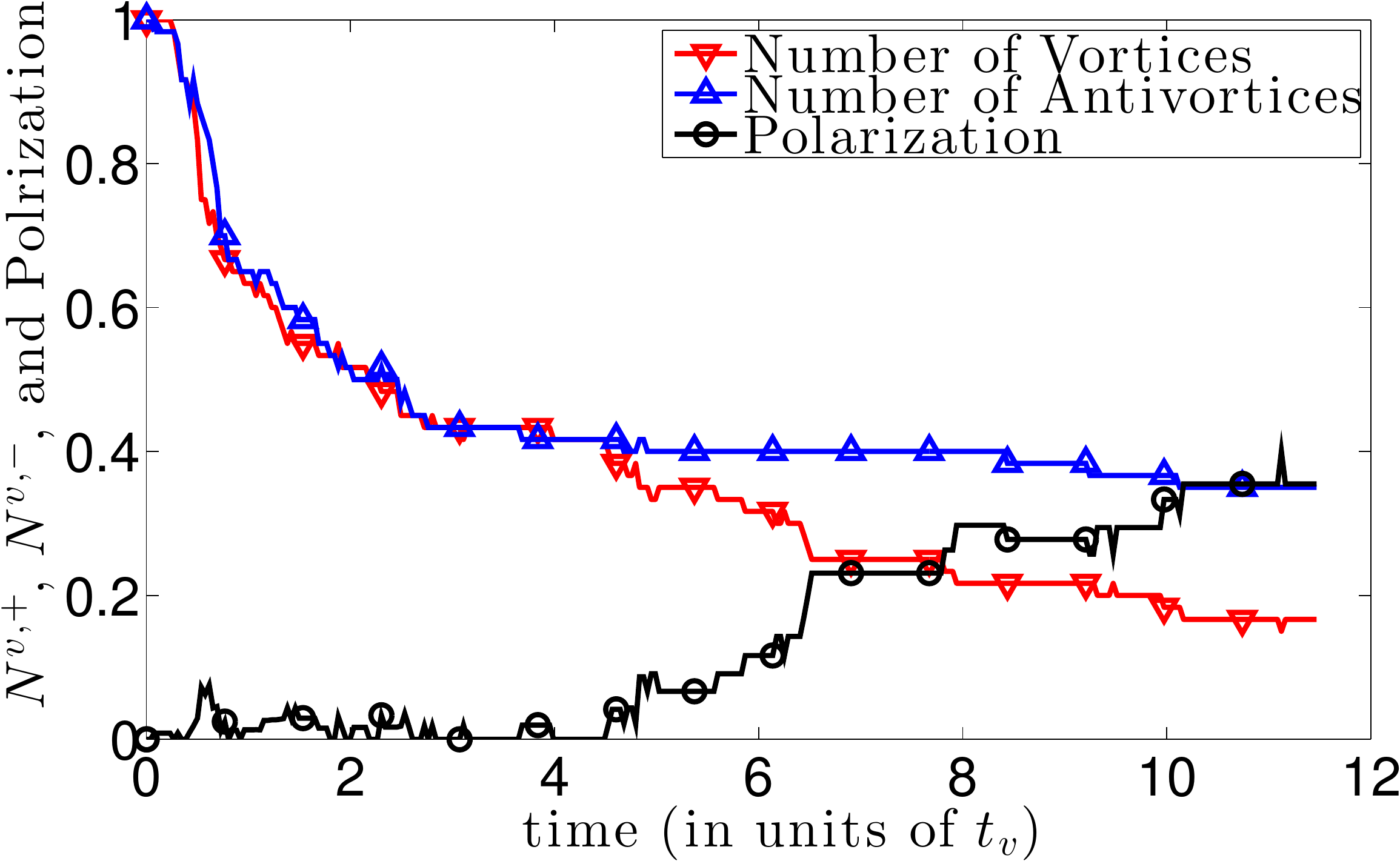}}
\subfigure[\label{fig_t0} Rectangle]{
\includegraphics[width=2.68in]{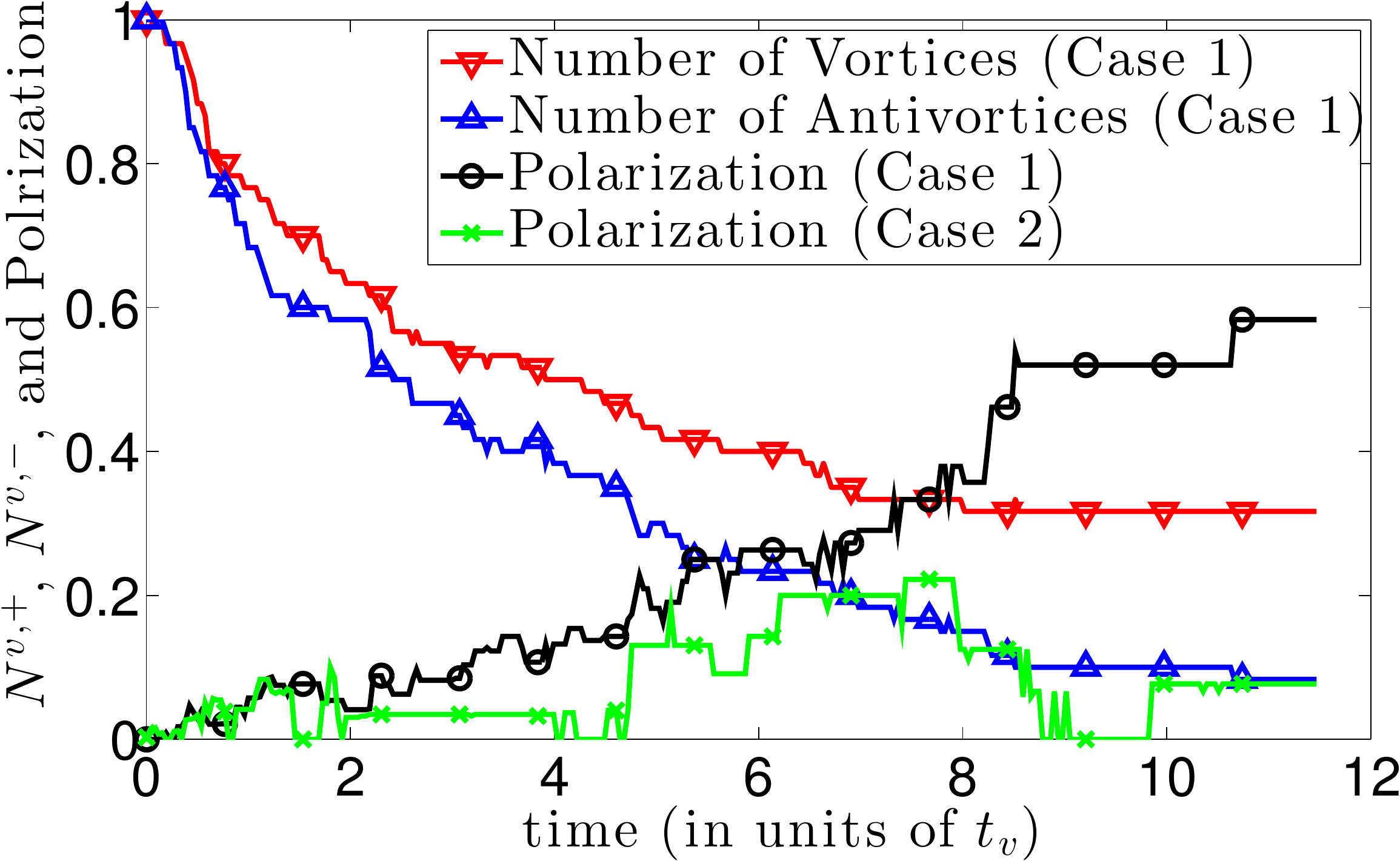}}
 \end{minipage}
 \vspace{-0.3cm}
  \caption{Time variation of total number of vortices and antivortices, and vortex polarization in GP simulations. \label{fig_Nvortices}}
\vspace{-0.4cm}
\end{figure}

The discrepancy between the dynamical runs and the mean field theory can be explained in terms of the polarization defined as ($N^{v,+} - N^{v,-})/(N^{v,+} + N^{v,-})$ where $N^{v,+}$ and $N^{v,-}$ are the number of vortices and antivortices. In Fig.\ \ref{fig_Nvortices}a, we observe that in the square, the increase in polarization coincides with when a monopole state emerges. In general, the polarization of the vortices can only change by single vortices annihilating at the boundaries of the domain \cite{Chomaz2015,Kwon2014}. Since the centrally located vortices of the monopole configuration are screened from the boundaries, this explains why this mean field solution favours the emergence of a polarized vortex state. In the rectangle, the perimeter of the domain is larger than in the square thus enhancing the effects of vortex annihilation at the boundaries. Moreover, since the monopole should emerge at earlier times in the rectangle as it has a lower entropy, these effects result in a polarized vortex state at earlier times as indicated by Fig.\ \ref{fig_Nvortices}b which causes the monopole to persist in the GP simulations. 

In rare realizations, the fluctuations can kick the system back from a polarized to a neutral state. Consequently, the dipole re-emerges in the rectangle at late times. In the particular example presented in Fig.\ \ref{fig_aveStream_PVGPrectangle}c (Case 2) corresponding to a realization with an initial total number of vortices $N^v=350$, and with $\tilde{g}=37346$ ($N$ reduced by a factor of $2.5$), we observe that the neutral polarization is restored at late times as seen in Fig.\ \ref{fig_Nvortices}b (see also \href{run:/Applications/VLC.app/Contents/MacOS/VLC ./supplemental/S3GPMonR.mp4}{S3GPMonR}). This explains why the dipole is recovered at late times in this GP simulation of the rectangle which coincides with the predictions of the mean-field theory that were obtained under the assumption that the vortex gas is not polarized.



\begin{figure}[t]
\centering
\hspace{-0.55cm}
 \begin{minipage}[b]{0.5\textwidth}
    \centering
    \subfigure[\label{fig_t0} Square]{
\includegraphics[width=1.68in]{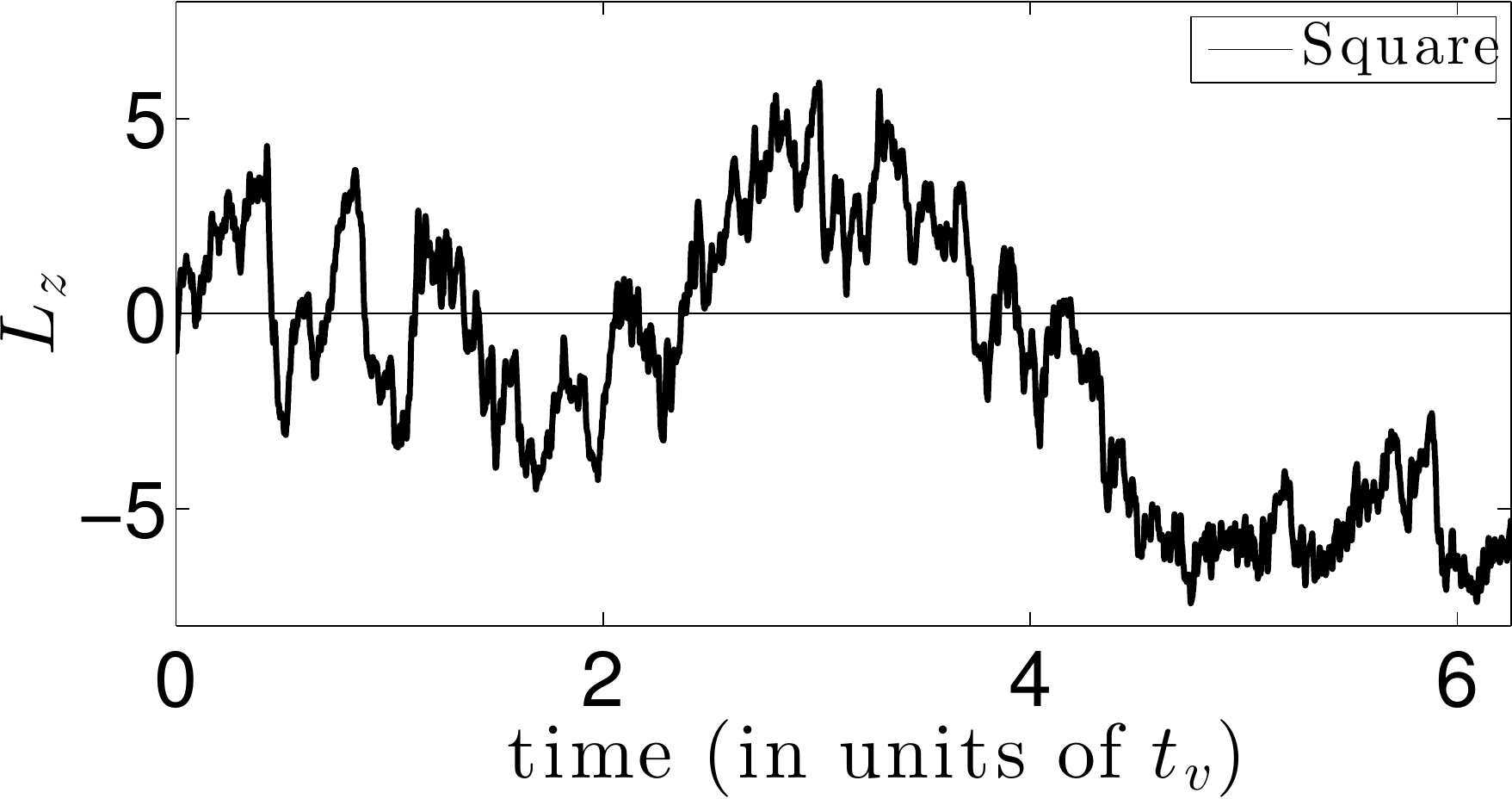}}
\addtocounter{subfigure}{-1}  
\subfigure[\label{fig_t0} Rectangle]{
\includegraphics[width=1.68in]{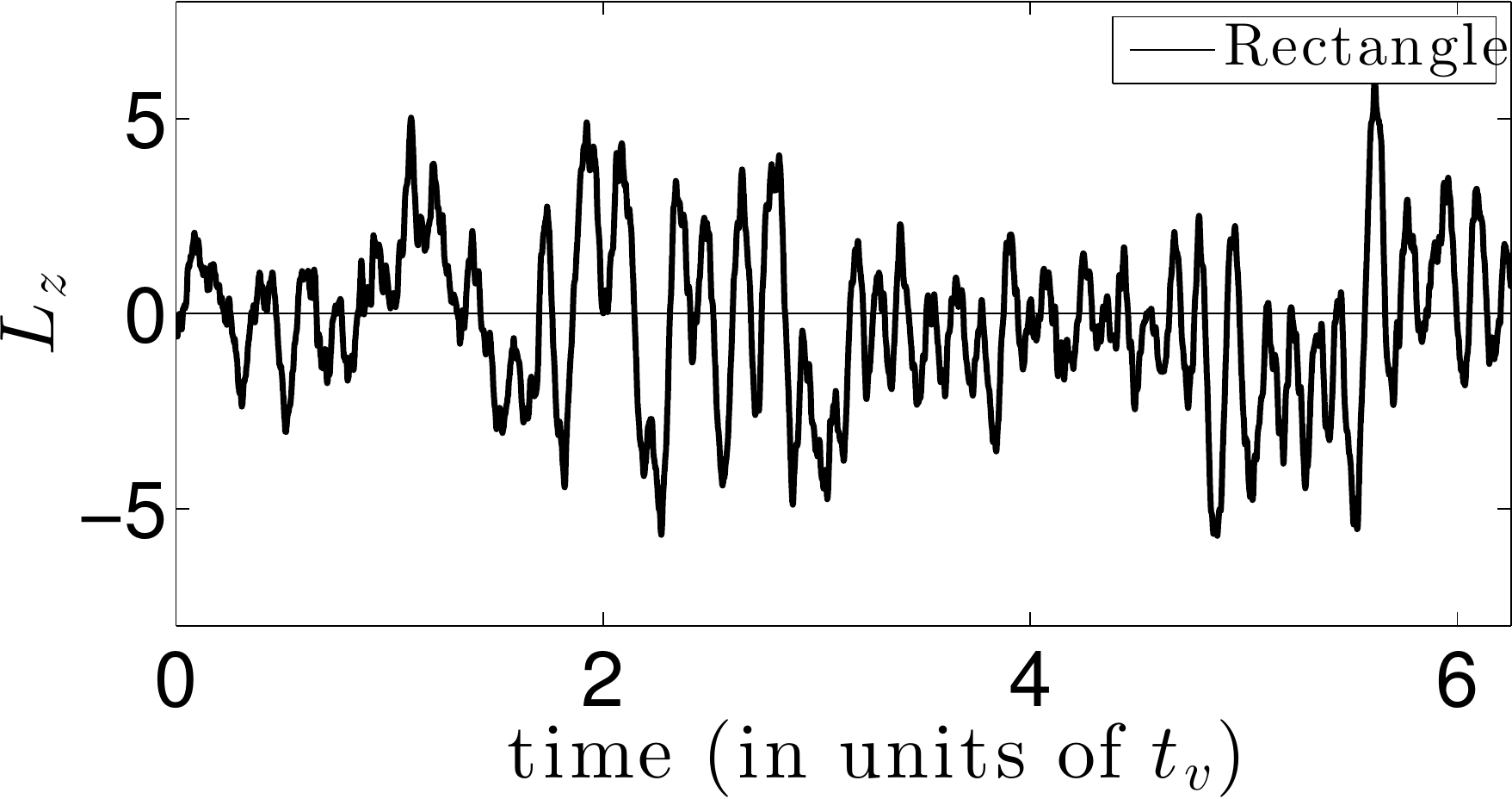}
}
 \end{minipage}
  \caption{Time variation of angular momenta in (a )square; (b) rectangular geometries for point vortex model.  \label{fig_AngMom}}
\end{figure}

The  coherent flows that emerge can be grouped into two types depending on whether or not these flows spontaneously acquire angular momentum, $L_z$. 
In Fig.\ \ref{fig_AngMom} we have presented the variation of $L_z$ with time in the point vortex model. As can be seen, a non-zero value of $L_z$ arises at late times in the square whereas it continues to fluctuate about zero for the rectangle. These qualitatively different coherent flows 
stand in contrast to the single coherent phase field in the case of the formation of a Bose-Einstein condensate in an atomic Bose gas. 

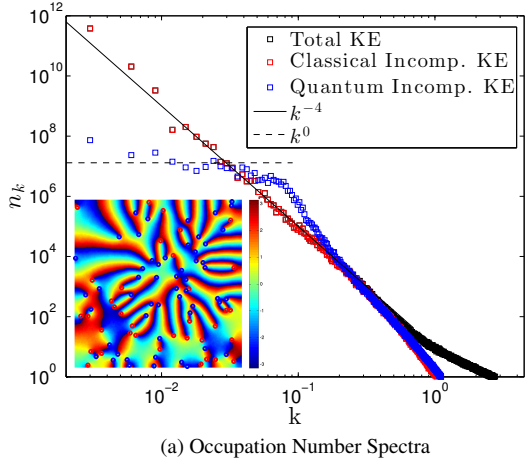
\begin{figure}[t]
\centering
 \begin{minipage}[b]{0.5\textwidth}
\subfigure[\label{fig_t0} Occupation Number Spectra]{
\sbox0{\includegraphics[width=2.7in]{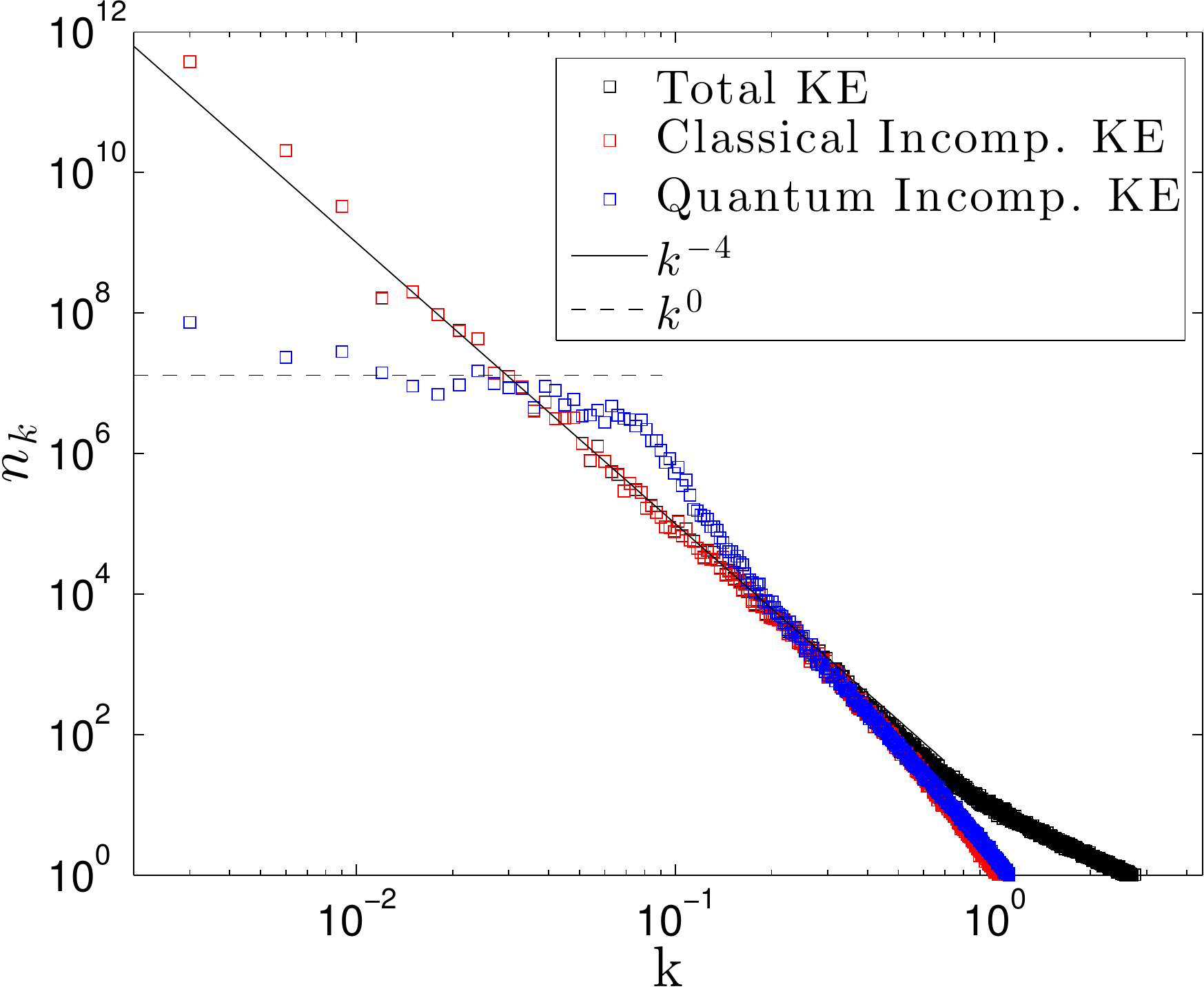}}
\sbox1{\includegraphics[width=1.in]{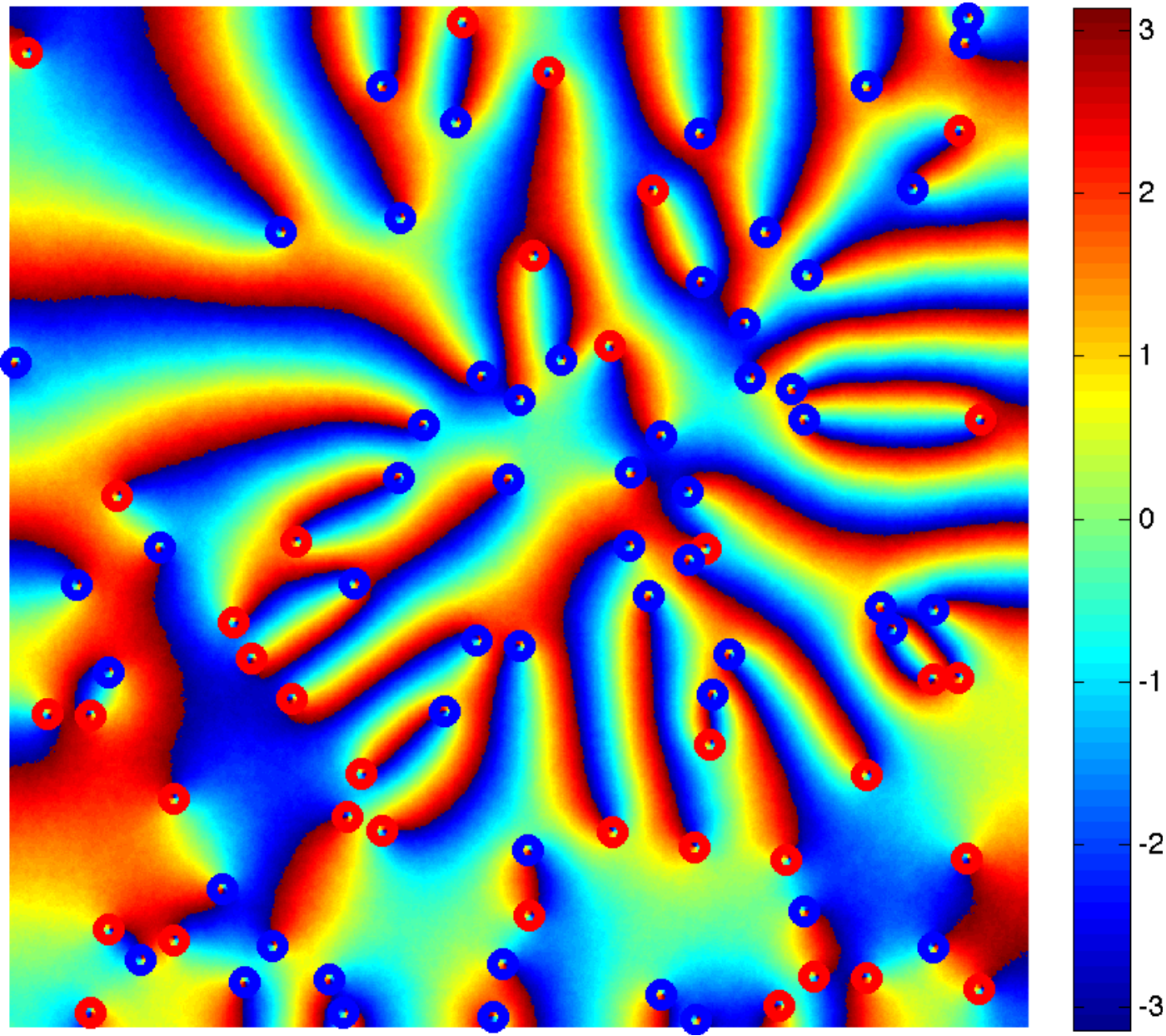}}
\begin{picture}(\wd0,\ht0)
\put(-0.5cm,0){\usebox0}
\put(\wd0 - \wd2 - 6.5cm,\ht0 - \ht2 - 4.875cm){\usebox{1}}
\end{picture}}
 \end{minipage}
 \vspace{-0.3cm}
  \caption{Kinetic energy spectra for GP simulation with $1025 \times 1025$ grid points with the interaction parameter set to $\tilde{g} = 282920$ and starting with an initial neutral configuration consisting of 360 vortices. Phase field shown in inset. \label{fig_Spectra}}
\vspace{-0.4cm}
\end{figure}

The coherent flow also has a clear signature in the occupation number spectrum of the wavefunction $\phi$. In Fig.\ \ref{fig_Spectra}, we have evaluated the angle averaged spectrum for a high resolution simulation of the GP equation in the square at late times after a coherent flow had emerged. Included are the occupation number densities corresponding to the classical and quantum incompressible kinetic energies (CIKE and QIKE respectively) \cite{Reeves2014} (see Appendix for further details of their definitions). We observe a clear difference between the two quantities at low wavenumbers. In fact, the two would coincide if the velocity field remains uncorrelated to the phase. Such a scenario corresponds to regimes considered in \cite{Nowak2011} where no coherent flow was identified. The emergence of the coherent flow in our case produces strong correlations between the phase and the velocity field and thus destroys the correspondence between the two quantities (see Appendix). The consequence is a characteristic flattening in the occupation number spectrum of the QIKE at low wavenumbers. 

This signature of the coherent flow can potentially be accessed from measurements of the momentum distribution of the Bose gas.
The flattening of the spectrum derived from the QIKE stands in contrast to the spectrum associated with the CIKE which slightly exceeds the $k^{-4}$ spectrum at low $k$ due to the effect of the spectral condensate. Indeed, as discussed in \cite{Reeves2014}, a clear signature of a spectral energy condensate appears in the CIKE spectrum only when the vortex gas is deep into the negative temperature regime whereas the QIKE spectrum is characterised by a $k^3$ power law in the infrared which corresponds to a flat spectrum in $n_k$. Therefore, the results of Fig.\ \ref{fig_Spectra} provide clear evidence that our simulations coincide with the spectral condensation of energy. 

We have shown that in 2D quantum fluids, large scale flows that characterise the spectral condensate in energy emerge as a result of 
a form of topological ordering that can occur at zero temperature and that is distinct from the BKT type. In contrast to previous works, by extending the Penrose-Onsager definition of a condensate to the streamfunction, we are able to define the spectral condensate in a directly analogous way to a BEC. We have shown how this definition can be combined with a theory for determining these coherent flows in a 2D quantum fluid. 
Our theory corroborates results of our numerical simulations that two different types of flows can emerge at large scales with each type distinguished by whether or not the flow spontaneously acquires angular momentum. Moreover, the simulations reveal that the mean-field predictions that are formally derived in the limit of a large number of vortices  appear to apply even when the total number of vortices is around 50-100. Since this is achievable in current experimental set-ups, our results suggest that it may be possible to confirm these predictions in future experiments.

The situation we describe draws a direct analogy with the scenario of a Bose condensed gas where a coherent mode coexists with thermally populated modes. In the same way that a separate treatment of the coherent and incoherent modes is essential in order to completely characterise the properties of a Bose condensed gas, we have presented a model that can explain the coherent spectral condensate of energy which allows us to characterise 2D quantum turbulence when combined with Kraichnan's theory \cite{Kraichnan1967,Kraichnan1980} of an inverse energy cascade.
The statistical theory that describes the spectral energy condensate in quantum fluids is also relevant 
to the phenomena of spectral condensation in 2D classical turbulence. Indeed the structures we have reported here for the decaying problem of quantum turbulence bear a striking similarity to flows observed experimentally even for forced 2D classical turbulence \cite{Xia2008,Xia2009,Xia2011}. 
This raises further open questions regarding the close relationship between the Boltzmann-Poisson theory and the Robert-Miller-Sommeria theory \cite{Robert1991a,Miller1990,Robert1991b} formulated for classical fluids and, in particular, which of the infinitely many Casimirs that ideal 2D classical fluids possess are needed in practice to correctly characterise the coherent flow.

\acknowledgements{The authors would like to thank Prof.\ G.~Esler, Prof.\ T.~Gasenzer, Prof.\ M.~Davis, Prof.\ A.~Bradley and Dr.\ D.~Proment for valuable discussions. HS acknowledges support for a Research Fellowship from the Leverhulme Trust under Grant R201540.}

\section{Appendix: Evaluating Spectral Distributions}

To evaluate the occupation number spectra, we note that the total kinetic energy is given by
\begin{eqnarray}
{E} = \int \big< |\grad \phi|^2 \big> {\rm d}x {\rm d}y = \int k^2 \big< |\tilde{a}({\bf k})|^2 \big> {\rm d}^2{\bf k} 
= 2\pi \int_0^{\infty} k^3 n({k}) {\rm d}k, \nonumber 
\label{eqn_KE}
\end{eqnarray}
where $n({k}) = 1/(2\pi)\int_0^{2\pi} \big<|\tilde{a}({\bf k})|^2\big> {\rm d} \theta_k$, and 
\begin{eqnarray}
\tilde{a}({\bf k}) \equiv \mathcal{F}[\psi({\bf r}] = \frac{1}{2\pi} \iint e^{-i{\bf k} \cdot {\bf r}} {\psi}({\bf r}) {\rm d}^2 {\bf r}.
\end{eqnarray}
Thereafter $\mathcal{F}[\cdot]$ will be used to denote the Fourier transformed quantity. It follows from the above that an energy spectrum with a $k^{-1}$ power law, consistent with the analysis in \cite{Nowak2011,Bradley2012}, corresponds to an occupation number spectrum with a $k^{-4}$ power law as observed at intermediate wavenumbers in our simulations.
To define the occupation number spectrum of the incompressible component of the kinetic energy, we first identify the hydrodynamic (H) and quantum pressure contributions (QP) to the kinetic energy given by
\begin{eqnarray}
{E}_{\text{H}}^{\text C} &=& \int \big< \rho({\bf r}) |{\bf v}({\bf r})|^2\big> {\rm d}^2{\bf r}
= \int_0^{\infty} {\rm d}k \int_0^{2\pi} \big<|\tilde{{\bf u}}({\bf k})|^2 \big> k {\rm d}\theta_k, \nonumber \\
{E}_{\text{QP}}^{\text C} &=& \int \big< |\grad \sqrt{\rho({\bf r})}|^2 \big> {\rm d}^2{\bf r}
\end{eqnarray}
where the velocity field is given by ${\bf v} = \grad \varphi({\bf r})$ and ${\bf u}({\bf r}) = \sqrt{\rho({\bf r})}{\bf v}({\bf r})$ corresponds to the the density-weighted velocity field and $\tilde{{\bf u}}({\bf k}) = \mathcal{F}({\bf u}({\bf r}))$. We use the superscript, C, to distinguish the classical definitions for the different components of the kinetic energy from the quantum definitions to the kinetic energy that include the phase dependent factors to be presented below. The incompressible and compressible contributions to the hydrodynamic component of the kinetic energy spectrum can be evaluated by utilizing the Helmholtz decomposition of the field ${\bf u}({\bf r})$. By writing ${\bf u} = {\bf u}^i + {\bf u}^c$ where $\grad \cdot {\bf u}^i = 0$ and $\nabla \times {\bf u}^c = {\bf 0}$, the classical incompressible kinetic energy (CIKE) is given by
\begin{eqnarray}
{E}^C_{\text{IH}} = \int_0^{\infty} {\rm d}k \int_0^{2\pi} \big< |\tilde{{\bf u}}^i({\bf k})|^2 \big> k {\rm d}\theta_k =  \int_0^{\infty} \mathcal{E}_{\text{CIKE}} {\rm d}k. \label{eqn_CIKE}
\end{eqnarray}
In relating the incompressible kinetic energy to the total kinetic energy spectrum as given by Eq.\ (\ref{eqn_KE}), it turns out that it is more natural to evaluate the incompressible kinetic energy from the modified definition given by
\begin{eqnarray}
{E}_{\text{IH}}^{Q} = \int \big< |\mathcal{F}[{\bf u}^i({\bf r}) e^{i \varphi ({\bf r})}]|^2\big> {\rm d}^2{\bf k} =  \int_0^{\infty} \mathcal{E}_{\text{QIKE}} {\rm d}k.
\label{eqn_QIKE}
\end{eqnarray}
We refer to this quantity as the Quantum Incompressible Kinetic Energy (QIKE). Similarly we can define 
\begin{eqnarray}
{E}_{\text{QP}}^{\text Q} &=& \int \big< |\mathcal{F}[\grad \sqrt{\rho({\bf r})} e^{i \varphi ({\bf r})}]|^2 \big> {\rm d}^2{\bf k}
\end{eqnarray}

As noted in \cite{Reeves2014}, this decomposition of the kinetic energy into different contributions provides genuine spectral energy densities that are locally additive in $k$ space in contrast to the classical spectra as given from Eq.\ (\ref{eqn_CIKE}). With these definitions, we can then define the respective occupation numbers corresponding to the classical and quantum kinetic energy spectra as $n_{\text{CIKE}}(k) = \mathcal{E}_{\text{CIKE}}/k^{3}$ and $n_{\text{QIKE}}(k) = \mathcal{E}_{\text{QIKE}}/k^{3}$.

Although the CIKE and the QIKE spectra generally differ, the two are equivalent provided the phase field $\varphi$ is uncorrelated to the density weighted incompressible velocity field ${\bf u}^i$. In particular, we have
\begin{eqnarray}
k^2 n(k) &=& \frac{1}{2\pi}\int_0^{2\pi} \left( \big< |\mathcal{F}[{\bf u}({\bf r}) e^{i \varphi ({\bf r})}]|^2({\bf k}) \big> \right. \nonumber \\
&& + \left. \big< |\mathcal{F}[\grad \sqrt{\rho({\bf r})} e^{i \varphi ({\bf r})}]|^2({\bf k}) \big> \right) \\ 
&& + \big< 2\mathcal{R} \left( i \mathcal{F}[{\bf u}({\bf r}) e^{i \varphi ({\bf r})}]({\bf k}) \mathcal{F}[\sqrt{\rho ({\bf r})} e^{i \varphi ({\bf r})}]^*({\bf k}) \right) \big> d\theta_k. \nonumber
\end{eqnarray}
where $\mathcal{R}(\cdot)$ denotes the real part. Now if the terms on the right-hand side are isotropic so they depend on $k$ only, then the third term vanishes \cite{Nowak2011}. The remaining terms can then be expressed as a convolution so that
\begin{eqnarray}
k^2 n(k) &=& \int_0^{2\pi} \left( \big< [\tilde{{\bf u}}*\mathcal{F}(e^{i \varphi ({\bf r})})](k)
[\tilde{{\bf u}}*\mathcal{F}(e^{i \varphi ({\bf r})})]^* \big> (k) \right. \\
&& \hspace{-0.7cm} + \left. \big< [\mathcal{F}(\grad \sqrt{\rho})*\mathcal{F}(e^{i \varphi ({\bf r})})](k)
[\mathcal{F}(\grad \sqrt{\rho})*\mathcal{F}(e^{i \varphi ({\bf r})})]^* (k) \big> \right) {\rm d}\theta_k. \nonumber 
\end{eqnarray} 
When the phase and velocity are uncorrelated, we can write the above in the form
\begin{eqnarray}
&& k^2 n(k) = \int {\rm d} {\bf p} \int {\rm d} {\bf q} \int_0^{2\pi} \big< [\tilde{{\bf u}}](|{\bf p}-\bk|)  [\tilde{{\bf u}}]^*(|{\bf q}-\bk|)
\big> {\rm d}\theta_k \nonumber \\
&\times& \int_0^{2\pi} \big<
[\mathcal{F}(e^{i \varphi ({\bf r})})]({\bf p}) [\mathcal{F}(e^{i \varphi ({\bf r})})]^*({\bf q}) \big> {\rm d}  \theta_k \nonumber \\
&+& \int {\rm d} {\bf p} \int {\rm d} {\bf q} \int_0^{2\pi} \big< [\mathcal{F}(\grad \sqrt{\rho})](|{\bf p}-\bk|)  [\mathcal{F}(\grad \sqrt{\rho})]^*(|{\bf q}-\bk|) \big> {\rm d}\theta_k \nonumber \\
&\times& \int_0^{2\pi} \big<
[\mathcal{F}(e^{i \varphi ({\bf r})})]({\bf p}) [\mathcal{F}(e^{i \varphi ({\bf r})})]^*({\bf q}) \big> {\rm d} \theta_k .
\end{eqnarray} 
If we furthermore assume that the phase is slowly varying so that $\mathcal{F}(e^{i \varphi ({\bf r})})$ is peaked at zero momentum,
we can approximate these terms by delta functions which consequently leads to the final expression
\begin{eqnarray}
k^2 n(k) \simeq  |\tilde{\bf u}|^2(k) + |\mathcal{F}(\grad \sqrt{\rho})|^2(k).
\end{eqnarray}
Hence, departures from this expression at low wavenumbers arise due to strong correlations developing between the flow and the phase when the large scale coherent flow emerges.

\bibliographystyle{aipnum4-1}
\bibliography{PRL}

\end{document}